\PassOptionsToPackage{square,comma,numbers,sort&compress}{natbib}
\documentclass[preprint,review,12pt]{elsarticle}
\usepackage{lineno}

\usepackage{graphicx}

\usepackage{tabularx} % Helpful for full text width tables.
\usepackage{graphicx} % Figures.
\usepackage{amssymb} % Math symbols.
\usepackage{soul} % Prevents underlined words/phrases from running into the margins.
\usepackage{mathtools}
\usepackage[version=4]{mhchem}
\usepackage{enumerate}
\usepackage{booktabs}
\usepackage{float}
\usepackage{indentfirst}
\usepackage{adjustbox}
\usepackage{etoolbox,siunitx}
\usepackage{threeparttable}

\usepackage{afterpage}
\usepackage{chngpage}
\usepackage{rotating}
\usepackage{hvfloat}
\usepackage{amssymb,amsmath,amsbsy,amsfonts}
\usepackage{bm}
\usepackage{bbm}
\usepackage{multirow}
\usepackage[bottom]{footmisc}
\usepackage[colorlinks=true,citecolor=red,linkcolor=blue]{hyperref}
\usepackage{placeins}
\usepackage{color}
\usepackage{comment}
\usepackage{setspace}
\usepackage{tcolorbox}
\tcbuselibrary{breakable}
\usepackage{pdflscape}

\tcbset{
  colback=gray!10,
  colframe=gray!50,
  fonttitle=\bfseries,
  boxrule=0.5pt
}

\usepackage[margin=1in]{geometry} % changing the margins

\journal{Peer Review}

\usepackage{footnote}
\usepackage{adjustbox}
\usepackage{lipsum}

\def \sdv {SD-v3.5-Medium }
\def \sdvv {SD-v3.5-Medium}

\begin{document}

\begin{frontmatter}

\title{NuclearDiffusion: Text-to-Image Foundation Models for Learning Nuclear Energy Concepts}
%\title{NuclearDiffusion: Learning Nuclear Concepts with Text-to-Image Foundation Models}

\author[a,b]{Mohammed I. Radaideh}
\author[d]{Jeremy Moon$^{**,}$}
\author[d]{Andre Gala-Garza$^{**,}$}
\author[d]{Emma Son}
\author[d]{Yug Shah}
\author[c,d]{Majdi I. Radaideh$^{*,}$}

\cortext[cor1]{Corresponding author: Mohammed I. Radaideh (malradai@umich.edu), Majdi I. Radaideh (radaideh@umich.edu)}
\cortext[cor2]{Authors (Jeremy Moon and Andre Gala-Garza) contributed equally}
%\fntext[label]{}

\address[a]{Department of Mechanical Engineering, University of Michigan, Ann Arbor, Michigan 48109, USA}
\address[b]{Michigan Institute for Computational Discovery and Engineering (MICDE), University of Michigan, Ann Arbor, Michigan 48109, USA}
\address[c]{Department of Nuclear Engineering and Radiological Sciences, University of Michigan, Ann Arbor, Michigan 48109, USA}
\address[d]{Department of Electrical Engineering and Computer Science, University of Michigan, Ann Arbor, Michigan 48109, USA}

\begin{abstract}
\small
Generative artificial intelligence (AI) has transformed text-to-image synthesis, yet its ability to represent specialized engineering domains remains largely unexplored. As an exmaple in nuclear engineering, general-purpose foundation models frequently generate physically incorrect or conceptually inconsistent images because they lack domain-specific knowledge. This work presents one of the first systematic studies of domain adaptation for nuclear text-to-image generation through fine-tuning of open-source diffusion models. We curate a dataset of 1,000 captioned nuclear energy images spanning reactors, fuel cycles, radiation, and related concepts, and use it to fine-tune three state-of-the-art open-source models: Stable Diffusion XL (SDXL), \sdvv, and the flow-matching Flux.1 model. Their performance is evaluated using both quantitative image-similarity metrics and qualitative expert assessment against the corresponding zero-shot models. Fine-tuning substantially improves the fidelity of SDXL, provides only limited gains for \sdvv, and yields no measurable improvement for Flux.1, demonstrating that adaptation effectiveness depends strongly on the underlying generative architecture rather than model scale alone. We further compare the fine-tuned models against three leading commercial systems—GPT-Image-2, Gemini-3.1-Flash-Image, and Midjourney. Although GPT-Image-2 and Gemini generate convincing images for broad nuclear concepts, they frequently fail on specialized engineering prompts, where the fine-tuned open-source models produce more accurate and technically consistent outputs. These results establish domain-specific fine-tuning as a practical pathway for developing trustworthy generative AI tools for domain-specific applications.

\end{abstract}

\setstretch{1.15}


\end{frontmatter}

%\noindent \textit{Disclosure: It is very important to mention that some of the prompts/examples used in this study might be offensive and harmful to certain backgrounds, groups, countries, and genders. These prompts are written for research and testing purposes only for this study and do not reflect the authors' beliefs.}


%%%%%%%%%%%%%%%%%%%%%%%%%%%%%%%%%%%%%%%%%%%%%
%%%%%%%%%%%%%%%%%%%%%%%%%%%%%%%%%%%%%%%%%%%%%
%%%%%%%%%%%%%%%%%%%%%%%%%%%%%%%%%%%%%%%%%%%%%
\excludecomment{toexclude}
 
%5   baranowski2017catalytic
%8   hackbarth2018recent
%9   burre2019production
%10 klokic2020investigations
%11 oestreich2018production
%12 peter2018towards
%13 ouda2018hybrid
%14 garcia2021methanol
%15 benajes2020potential
%16 omari2019potential
%19 popp2019potentials
%21wang2020numerical

%%%%%%%%%%%%%%%%%%%%%%%%%%%%%%%%%%%%%%%%%%%%%
\section{Introduction}
\setstretch{1.15}
\label{intro}

Artificial Intelligence (AI) is rapidly transforming engineering by accelerating scientific discovery, automation, design optimization, and decision-making across a wide range of disciplines \cite{arhouni2025artificial}. Among the most significant recent advances are generative AI models capable of synthesizing realistic images from natural language descriptions. In particular, diffusion-based models have revolutionized text-to-image generation by producing high-quality images from textual prompts \cite{bansal2024revolutionizing}, while newer flow-matching architectures have further improved image quality and generation efficiency \cite{schusterbauer2025diff2flow}. These foundation models are increasingly being adopted for applications such as education, scientific visualization, conceptual design, and digital content generation \cite{sanseviero2024hands}. Such capabilities have sparked growing interest in leveraging generative AI to assist domain experts by rapidly creating visual representations that would otherwise require substantial manual effort. However, the successful application of these models depends critically on their ability to understand the specialized concepts, terminology, and visual characteristics of the target domain.

Nuclear engineering presents one of the most challenging domains for text-to-image generation because the generated images must be not only visually realistic but also technically and scientifically accurate. Unlike common objects found in large public image datasets, nuclear concepts such as fuel assemblies, reactor cores, containment structures, spent fuel, fusion systems, and advanced reactor designs are highly specialized and relatively scarce in publicly available training data. Consequently, state-of-the-art zero-shot text-to-image models frequently generate conceptually incorrect or physically unrealistic nuclear imagery, limiting their usefulness for education, communication, and engineering applications. Although AI has become increasingly prevalent in nuclear engineering, early applications primarily relied on conventional machine learning techniques \cite{fernandez2017nuclear,radaideh2020surrogate}, with more recent work expanding to deep learning \cite{bae2020deep,radaideh2020neural}, reinforcement learning \cite{lee2020algorithm,radaideh2021physics}, and large language models \cite{kwon2024sentiment,lee2025large}. In contrast, comparatively little attention has been devoted to adapting modern generative image models for nuclear-specific applications. This work addresses this gap by fine-tuning state-of-the-art open-source text-to-image foundation models on a curated dataset of nuclear imagery and systematically evaluating whether domain-specific fine-tuning can substantially improve the conceptual fidelity and technical accuracy of generated nuclear images compared with leading commercial image-generation systems.

To address this gap, this work develops and systematically evaluates domain-adapted text-to-image foundation models for nuclear engineering. Rather than relying on zero-shot inference from general-purpose pretrained models, we investigate whether domain-specific fine-tuning can enable modern generative AI models to accurately capture specialized nuclear concepts despite the scarcity of high-quality training imagery. The resulting framework demonstrates how foundation models can be adapted to safety-critical engineering domains where conceptual correctness is as important as visual realism. The main contributions of this work are summarized as follows:

\begin{itemize}
\item[1.] We present one of the first comprehensive studies on fine-tuning state-of-the-art open-source text-to-image foundation models, including Stable Diffusion and Flux, for nuclear engineering. The proposed models learn a broad range of nuclear concepts, including reactor technologies, fuel assemblies, uranium fuel, fusion systems, spent fuel management, and advanced reactor applications, enabling the generation of technically meaningful nuclear imagery from natural language prompts.

\item[2.] We introduce a curated dataset of 1,000 captioned nuclear images covering diverse reactor technologies and nuclear engineering concepts. The dataset provides a foundation for future research in domain-specific generative AI, including text-to-image, text-to-video, and benchmarking methodologies for evaluating technical image generation.

\item[3.] We perform a comprehensive evaluation of the fine-tuned models using both quantitative metrics and expert qualitative assessment to measure improvements in visual realism, conceptual fidelity, and technical accuracy for nuclear image generation.

\item[4.] We benchmark the fine-tuned models against leading commercial text-to-image systems, including GPT-Image-2, Gemini-3.1-Flash-Images, and Midjourney, demonstrating the advantages and limitations of general-purpose foundation models compared with domain-adapted models for nuclear engineering applications.

\end{itemize}

The remainder of this paper is organized as follows. Section~\ref{sec:lit_review} reviews the existing literature and related work on machine learning applications and text-to-image generation, with particular emphasis on nuclear engineering. Section~\ref{sec:method} describes the nuclear image collection process, the automated Python pipeline for extracting images from publications, the dataset preparation procedure, and the text-to-image models used for fine-tuning. Section~\ref{sec:eva} presents the quantitative evaluation metrics and qualitative assessment methodology. Section~\ref{sec:forward_res} discusses the experimental results, comparisons with commercial models, and additional analyses. Finally, Section~\ref{sec:conc} concludes the paper with limitations and future research directions.

\section{Literature Reivew \& Related Work}
\label{sec:lit_review}
\subsection{Evolution of Artificial Intelligence in Nuclear Engineering}
\label{sec:ai_nuclear}

AI has become an important computational tool across the nuclear energy lifecycle, supporting reactor modeling, design optimization, safety analysis, monitoring, and autonomous operation. Recent reviews show growing use of machine learning for fault diagnosis, equipment monitoring, predictive maintenance, reactor control, and decision support \cite{jendoubi2024survey}, although limited plant data, regulatory uncertainty, interpretability, and deployment readiness remain major barriers \cite{hall2024barriers}. Deep neural networks have been widely used as accelerated surrogates for computationally expensive simulations, including uncertainty-aware energy modeling \cite{radaideh2019combining} and high-dimensional boiling water reactor neutronics \cite{saleem2020application}. Automated machine-learning platforms have further streamlined model development and comparison \cite{myers2025pymaise}, while interpretable architectures based on Kolmogorov--Arnold Networks and symbolic regression have been explored to improve model transparency \cite{panczyk2025opening}. AI has also enabled large-scale optimization of reactor fuel and control systems through neuroevolution \cite{radaideh2021large} and hybrid swarm and evolutionary algorithms \cite{price2022multiobjective,radaideh2022pesa}, marking a shift from passive prediction toward active engineering design.

Deep reinforcement learning has advanced this shift by enabling multistep control strategies learned directly from reactor simulators. Applications include automated power ascension \cite{lee2020autonomous}, plant heat-up \cite{park2022control}, and coordinated multi-objective operation \cite{bae2023multiobjective}. A recent review identified reactor startup, load following, coordinated control, and emergency response as promising applications, while noting unresolved challenges in safety assurance, interpretability, and physical deployment \cite{gong2024possibilities}. In microreactors, reinforcement learning has been applied to transient control and load following \cite{tunkle2025nuclear}, as well as criticality search and power shaping \cite{radaideh2025multistep}. In parallel, AI-enabled digital twins are integrating measurements, physics-based simulations, and data-driven models. Recent developments include physics-informed active learning \cite{nabila2026physics}, uncertainty-aware variational twins \cite{burnett2026variational}, graph-neural-network whole-system reactor twins \cite{liu2025digital}, and neural-operator virtual sensing for thermal-hydraulic fields \cite{hossain2025virtual}. However, most nuclear digital twins remain at low or intermediate maturity because of persistent challenges in validation, data integration, cybersecurity, and real-time synchronization \cite{huang2025shaping}.

Large Language Models (LLMs) and generative AI are now extending nuclear AI beyond prediction and control toward language understanding, knowledge retrieval, human--AI interaction, and content generation. Early applications include nuclear sentiment classification \cite{kwon2024sentiment,kwon2024using}, analysis of paraphrasing, sarcasm, emojis, and other linguistic nuances \cite{bhargava2026impact}, and fairness and social-bias assessment in nuclear communication \cite{radaideh2025fairness}. LLM agents have also been connected to reactor simulators to interpret plant information, retrieve operating guidance, and support reactor-operation tasks \cite{lee2025agent}. Generative AI is further being explored for synthetic engineering-data development, with generative adversarial networks used to augment limited datasets and improve data efficiency \cite{nabila2025data}. Together, these advances show a progression from models that predict predefined outputs toward foundation models capable of generating data, language, recommendations, and other technical content.

Despite this rapid evolution, generative AI research in nuclear engineering remains concentrated on structured numerical data and natural language. Comparatively little work has examined visual foundation models or determined whether general-purpose text-to-image models can learn the specialized components, geometries, and physical relationships that characterize nuclear systems. This gap limits the use of generative AI for technically accurate visualization, education, scientific communication, and conceptual design, motivating the domain-specific adaptation investigated in this study.

\subsection{Text-to-Image Generative AI in Specialized Scientific and Engineering Domains}
\label{sec:t2i_domains}

Text-to-image generative AI has expanded from general-purpose content creation into architecture, medicine, education, and other specialized domains. In architecture, these models can rapidly generate alternative concepts and styles, but they often fail to reproduce region-specific features or technically coherent designs. For example, Midjourney, DALL-E~2, and Stable Diffusion were unable to consistently represent Emirati architectural elements \cite{ibrahim2025comparative}, while architecture-focused generative models produced visually interesting but spatially incoherent concepts \cite{horvath2024ai}. Similar studies suggest that such tools are more useful for early-stage ideation than for technically validated final designs \cite{liu2023application,karadaug2025new}. In medicine, domain-specific models have shown stronger performance. MedM2G improved the generation of X-ray, MRI, and CT images compared with GAN- and diffusion-based baselines \cite{zhan2024medm2g}, while a latent diffusion model combined with CLIP achieved improved semantic alignment and image-quality metrics \cite{kim2024controllable}. However, expert evaluations of DALL-E~3 images of congenital heart diseases revealed substantial anatomical inaccuracies despite their visual appeal \cite{temsah2024art}.

Text-to-image models have also supported design education and creative exploration by enabling rapid visualization of alternative styles and concepts \cite{liao2023study,derevyanko2023comparative}. At the same time, studies have identified demographic biases in generated depictions of academics and medical professionals, although the direction and magnitude of these biases vary across models and prompts \cite{currie2025gender,currie2024gender,bui2026portrayal}. Overall, the literature shows that visually convincing images may still contain technical, anatomical, contextual, or representational errors. Performance is strongly influenced by the availability of domain-specific data, and conventional metrics may not detect inaccuracies that are evident to subject-matter experts. These findings highlight the need for domain-specific adaptation and expert-centered evaluation in technically sensitive applications.

\subsection{Generative Visual AI in Nuclear Engineering}
\label{sec:nuclear_genai}

Generative visual AI remains largely unexplored in nuclear engineering, where most AI applications have focused on prediction, monitoring, anomaly detection, optimization, control, and decision support. Existing work on synthetic nuclear imagery can be grouped into three limited directions. First, Branikas et al. used a CycleGAN to generate crack images for data augmentation in visual inspection, improving crack-detection performance when real defect data were scarce \cite{branikas2023novel}. Second, Knotek et al. used explicitly defined three-dimensional fuel-assembly scenes and the Mitsuba~3 renderer to generate synthetic images and videos under controlled defect and inspection conditions \cite{knotek2025simulating}. Although useful for inspection and algorithm validation, both approaches are restricted to narrow tasks and require either existing reference images or manually constructed scenes.

The most directly relevant work evaluated general-purpose text-to-image models for nuclear-energy prompts. Our previous study compared 20 systems and found that DALL-E, DreamStudio, Craiyon, and related models could generate plausible images for broad prompts but frequently failed to represent technically specialized concepts such as reactor cores and fuel assemblies \cite{joynt2024comparative}. These findings show that zero-shot inference and prompt engineering alone are insufficient for reliable nuclear image generation. The literature therefore lacks a diverse captioned nuclear image dataset, systematic fine-tuning of modern diffusion and flow-matching models, evaluation methods that distinguish visual realism from technical correctness, and direct benchmarking of domain-adapted models against commercial systems. The present study addresses these gaps through nuclear-specific dataset development, model fine-tuning, and complementary quantitative and expert qualitative evaluation.

\section{Methodology}
\label{sec:method}

\subsection{Image Collection}

Images and their captions are needed to fine-tune text-to-image models. Accordingly, we searched for real images of nuclear energy and nuclear reactors in comprehensive references, including handbooks, nuclear energy encyclopedias, publications, and websites that publish news related to nuclear energy. This part was one of the most challenging parts of the study.

\subsubsection{Nuclear Energy Books}

The nuclear energy books provide the most reliable information on nuclear energy. However, books on nuclear physics and reactor theory contain few real images, and most of their figures are not useful, such as plots and flow charts. Handbooks and encyclopedias contain many more useful images, such as real-world images related to nuclear energy and schematics that illustrate concepts. However, some handbooks had only black-and-white images and were therefore excluded. References considered are Encyclopedia of Nuclear Energy \cite{greenspan2021encyclo}, Handbook of Nuclear Engineering Volumes 1-5 \cite{dan2010handbook}, and Handbook of Generation IV: Nuclear Reactors \cite{pioro2010hanbook}. Images and their captions were automatically extracted using Python, as will be explained shortly, and then manually checked. This manual check was necessary to filter out non-meaningful images that the automated procedure missed, such as 2D and 3D plots, contour plots, and flow charts. Filtering out such images reduces the number of usable images to 348. Other images unrelated directly to nuclear energy, such as non-nuclear steam generators, were also excluded.

\subsubsection{Nuclear Energy/Engineering Publications}
Articles published in nuclear energy and engineering journals can be a source of abundant images related to nuclear energy. Accordingly, our team downloaded 11,600 publications from Elsevier's nuclear energy/engineering journals: (1) Nuclear Engineering and Design, (2) Nuclear Engineering and Technology, (3) Annals of Nuclear Energy, and (4) Progress in Nuclear Energy. The manual extraction of images from around 11,600 PDFs, including books and encyclopedias, is cumbersome and impractical. Therefore, our team developed a Python pipeline to automate the process and filter out unwanted images, such as plots, bar charts, and contour plots. The pipeline consists of three stages:

\begin{itemize}
    \item [1.] \textbf{Zero-shot classification using CLIP-ViT-L/14 Model}. At this stage, images are extracted from the PDFs, and text-image alignment is assessed using the ViT-L/14 model \cite{radford2021learning} with a set of prompts that either keep or reject the images. The sets of prompts to keep and reject images are reported in the supplementary material. As an additional filter, the images are compared with reference images related to nuclear energy. \textit{The reference images used are the ones extracted from handbooks and the encyclopedia.}
    
    \item [2.] \textbf{Images Filtration using GPT-4.1-nano.} The remaining images from stage 1 are then passed to GPT-4.1-nano to filter out the unwanted images and keep only nuclear schematics. The prompt used is quite long and reported in \ref{sec:appA} for reference.

    \item[3.] \textbf{Images Filtration using GPT-5-nano}. The remaining images from stages 1 and 2 are then passed to GPT-5-nano, where the prompt is identical to that used for GPT-4.1-nano and reported in \ref{sec:appA} for reference.
\end{itemize}

The first stage (least accurate) of the pipeline extracted 41,342 captioned images from 11,600 PDFs. The second stage (GPT-4.1-nano) significantly reduced the number to 13,453 images, and the third stage (GPT-5-nano) further reduced it to 3,441 images. We then manually inspected the 3,441 remaining images and still noticed a significant number of non-useful images, such as plots, diagrams, incomplete images, and images with non-meaningful captions like ``the experimental setup''. Therefore, our team filtered out the non-useful ones, leaving only 119.

\subsubsection{Nuclear Energy News Websites}

The very limited number of useful images extracted from publications, compared to the number of processed PDFs, forced us to seek a better alternative: websites that publish daily news on nuclear energy. Such websites provide not only real images but also captions, though not for all images. The most active websites include the American Nuclear Society (ans.org), World Nuclear News (world-nuclear-news.org), Nuclear Engineering International (neimagazine.com), NUCNET (nucnet.org, an access request is required), and POWER Magazine (powermag.com). The images and captions were extracted manually with caution to avoid many duplicates, as these websites may publish articles on the same subject using the same images and captions. We extracted 533 images from these websites. We also collected 10 images from the British Science Museum website (sciencemuseum.org.uk), bringing the total to 1,000 captioned images. Table \ref{tab:res} summarizes the different sources we looked at and the number of recovered images. 

\begin{table}[h!]
  \centering
  \caption{The three Nuclear Energy (NE) image resources we used and the number of images collected from each one. The total sums to 1,000 captioned images.}
    \begin{tabular}{lc}
    \toprule
    Source          & Number of Images Collected \\
    \midrule
    NE Books        & 348 \\
    NE Publications & 119 \\
    NE Websites     & 533 \\
    \bottomrule
    \end{tabular}%
  \label{tab:res}%
\end{table}%

To our knowledge, there is no recommended number of images to fine-tune Stable Diffusion (SD) or Flux models to produce high-quality images. The best we can quote is the minimum specified in \cite{sanseviero2024hands} for SD full-model fine-tuning: 500 images. Since our dataset exceeds this minimum recommendation by approximately a factor of two, we believe it provides a sufficiently robust foundation for the initial fine-tuning and evaluation of the models. It should be noted that, unlike natural-image domains, nuclear engineering is characterized by a limited availability of publicly accessible, high-quality annotated images. Consequently, large-scale dataset expansion is challenging, and the most practical approach to increasing dataset diversity is through carefully designed augmentation techniques or the generation of controlled variations of existing reference images while preserving their underlying physical and semantic content.

\subsection{Image Extraction and Preprocessing}

\begin{figure}[h!]
    \centering
    \includegraphics[width=0.4\linewidth]{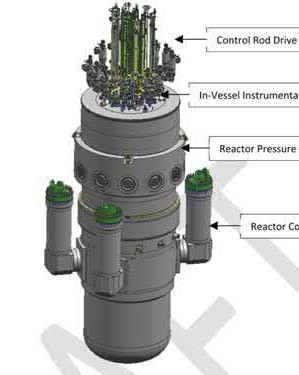} \hspace{0.5cm} \includegraphics[width=0.55\linewidth]{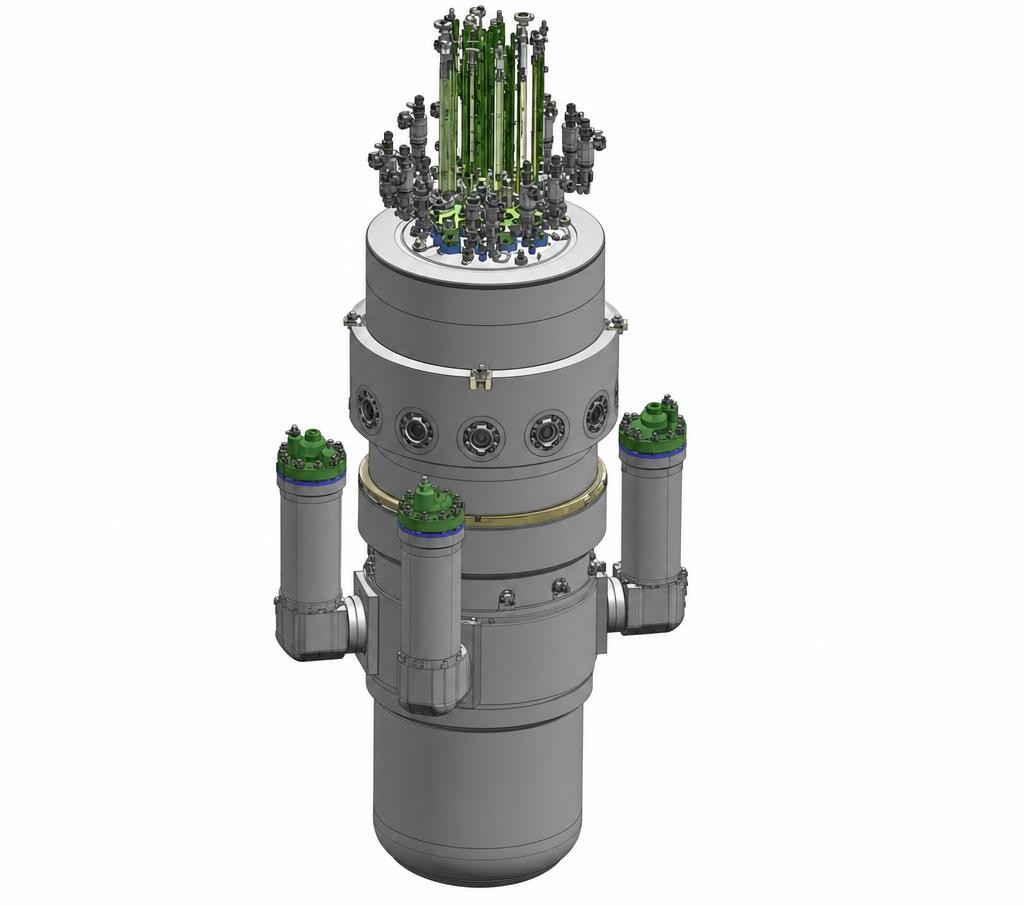}
    \caption{An example of an image preprocessed using the GPT-image-2.0 model, where the image on the left is the original image and the image on the right is after preprocessing. The image represents a small modular reactor configuration of RITM-200 developed by OKBM Afrikantov.}
    \label{fig:preproc}
\end{figure}

Image sizes (resolution) for Stable Diffusion models range between (512$\times$512) and (1024$\times$1024). However, a significant number of the extracted images are smaller than 512 pixels. Therefore, we used the latest GPT image model, GPT-Image-2, to upscale all images to 1024$\times$1024 and improve image quality. We also asked the GPT model to remove all text and arrows from the images because we noticed that the SD and Flux.1 models were unable to generate images with readable text, even after fine-tuning. The prompt used for GPT-Image-2 is ``\textit{Upscale this image to 1024$\times$1024 with improved quality. Remove all: text, arrows, labels, annotations, callouts. Preserve the entire original frame and object. Do not crop anything. Keep the image realistic and clean.}''. GPT-Image-2 model performed very well in the vast majority of the images; Figure \ref{fig:preproc} shows samples before (left) and after (right) using GPT-Image-2 to modify one image, where the image was upscaled to (1024$\times$1024) from (299$\times$375), its quality was increased, and all arrows and text were removed.

OpenAI Application Programming Interface (API) was used to automate image processing in Python. Processing the 1,000-image dataset incurred a cost of approximately \$200, which was affordable for the scope of this study. However, the cost is expected to scale approximately linearly with dataset size, making large-scale fine-tuning considerably more expensive. Both the computational cost and inference time are also influenced by the selected output quality. In this work, we used the ``high'' quality setting—the highest-fidelity and most computationally expensive option—to maximize image quality. Under these settings, processing the entire dataset required several days to complete.

\subsection{Text-to-Image Models Used and Fine-Tuning Setup}

Stable Diffusion (SD) models are latent diffusion models, meaning they compress the original image into a latent space where the noising-denoising diffusion process occurs. SD models use a variational Autoencoder (VAE) to compress the image into a latent space. Then, random Gaussian noise is added to the image. A neural network called UNet learns to predict noise, i.e., to remove it, in a number of steps, called time steps. In parallel, at each time step, this UNet also takes the image caption encoded by a text encoder as input conditioning, alongside the noisy image latents \cite{sanseviero2024hands}. SD models are open-source available in HuggingFace, fine-tunable, and usable for inference. In this study, we will use two SD models: SDXL and SD-v3.5-medium. SDXL can be fine-tuned using the Python package diffusers \cite{von-platen-etal-2022-diffusers}, but cannot be used to fine-tune \sdvv. The developers (StabilityAI) recommend the Python package SimpleTuner \cite{simpleTuner} for \sdv full model fine-tuning, as it uses the diffusers package. 

Besides the SD diffusion models, we also leveraged the open-source Flux.1 model available on HuggingFace, which operates on the flow-matching concept rather than diffusion denoising. In flow-matching, a transformer learns to predict a velocity field to obtain the real image from given noise, rather than adding the noise to the image and learning to remove it \cite{labs2025flux}. Flux.1 has 12 billion parameters and requires massive GPU memory for full model fine-tuning, which the authors currently do not have. Therefore, we used a Python procedure to fine-tune Flux.1 using Low-Rank Adaptation (LoRA) \cite{hu2022lora} provided by a third-party package, AI-Toolkit \cite{ai_toolkit}. In LoRA training, a subset of the 12 billion parameters is trained while the rest are frozen. This significantly reduces the required GPU VRAM for fine-tuning Flux.1, making it feasible, but it may negatively affect the model's performance.

For fine-tuning SD models (SDXL, SD-v3.5-Medium), we used 150 epochs. For diffusers, we could only specify the number of epochs, and the number of training steps is calculated accordingly. On the other hand, SimpleTuner is more flexible, and we could specify either the number of epochs or the number of training steps. For SD models, our hyperparameter setting ensured 150 epochs. For Flux.1, we could only specify the number of training steps. Therefore, we will present the quantitative results as a function of the number of training steps. The number of training steps for the SD models is 37,500. Note that the number of epochs or training steps depends on other hyperparameters, including batch size and the number of gradient accumulation steps. The batch size for SDXL is 1, and the number of gradient accumulation steps is 4. For SD-v3.5-Medium, the batch size used is 8, and the number of gradient accumulation steps is 1. The learning rate for the two SD models is 3$\times10^{-5}$. More detailed hyperparameter settings are reported in the \textbf{Supplementary Materials} (as a spreadsheet) and are used in the training packages: diffusers (SDXL) and SimpleTuner (SD-v3.5-Medium). 

For Flux.1, we used the 2$\times$ maximum value recommended by the developers: 2$\times$4,000 = 8,000 training steps. Fine-tuning the full Flux.1 model is not feasible on our available GPUs; therefore, we used Low-Rank Adaptation (LoRA) with a high rank of 256 to capture the complex details in our nuclear energy images. We used batch sizes of 1, with the choice depending mainly on the available GPU VRAM, as larger batch sizes can eventually lead to out-of-memory errors. The learning rate is 3$\times10^{-5}$. As with SD models, additional settings are available in the spreadsheet in \textbf{Supplementary Materials} and are compatible with the fine-tuning code provided by AI-Toolkit.

All three models were fine-tuned using an NVIDIA RTX 6000 Ada GPU with 48 GB of VRAM. It should be noted that while we have three other similar GPUs, all three fine-tuning packages (diffusers, SimpleTuner, AI-Toolkit) do not distribute models across multiple GPUs; instead, they use them to parallelize fine-tuning by dividing the number of training images among them. The fine-tuning time varies: around 32 hours for SDXL, around 60 hours for SD-v3.5-Medium, and around 9 hours for Flux.1. For inference, 50 steps were used to generate images for zero-shot models, and 15 checkpoints were saved after training at [2,500, 5,000, ..., 37,500] steps for SDXL and SD-v3.5-Medium. For Flux.1, 16 checkpoints were saved at training steps [500, 1000, ..., 8000]. We are using 300 testing prompts; the inference time for 15-16 checkpoints $\times$ 300 prompts was also quite significant and extended to days for Flux.1 (40-45 seconds per image).

\section{Evaluation Strategy}
\label{sec:eva}

\subsection{Quantitative Assessment Strategy}
\label{subsec:quant}

The evaluation of the images generated by SD models can be qualitative and quantitative. The qualitative assessment is the ideal approach, relying on human judgment and practical for a limited number of images. However, the cost of human involvement becomes high and impractical for large sets of generations. Therefore, we will use two quantitative metrics to identify the best checkpoint and compare it with a qualitative assessment of the best models identified by these metrics to further validate them. 

There are several quantitative metrics for comparing generated images with references; a common one is the Fr\'echet Inception Distance (FID) \cite{lee2023fidgan}. However, FID is biased with small sample sizes and requires up to 20,000 images to yield consistent values \cite{binkowski2018demystifying,jayasumana2024rethinking}. Accordingly, unbiased metrics based on Maximum Mean Discrepancy (MMD) distance, such as Kernel Inception Distance (KID) \cite{binkowski2018demystifying} and Contrastive Language–Image Pretraining (CLIP)-MMD (CMMD) \cite{jayasumana2024rethinking}, were introduced. KID and CMMD are not affected by sample size and use kernel functions rather than assuming normal distributions, as in FID. The kernel function is required to calculate the MMD distance between the real and generated image embeddings. CMMD is more recent than KID and uses a CLIP image embedding model trained on 400 million text-image pairs, whereas the KID image embedding model was trained on only 1 million images \cite{jayasumana2024rethinking}. In this study, we will calculate both CMMD and KID for 300 images generated using 300 prompts randomly selected from the 1,000 prompts used for fine-tuning. For both KID and CMMD, the \textbf{lower} the metric value, the better the similarity between the generated images and the reference ones.

It should be noted that several other metrics provide numerical scores for generated images without requiring real images for comparison \cite{tian2025quality}, as in KID and CMMD. Accordingly, testing prompts that \textit{differ significantly} from the original fine-tuning prompts can be assessed. Among such metrics is the infamous Contrastive Language–Image Pretraining (CLIP) score \cite{hessel2021clipscore}, which assesses text-image alignment for text-to-image models. However, such metrics rely on image embedding models that are likely to fail with unfamiliar images, such as those related to nuclear energy that we collected in our study. Therefore, some of these embedding models must be trained on nuclear energy images to make these metrics more tailored to assess synthetic nuclear energy imagery. Nevertheless, this is beyond the scope of this work and will be explored in future work. Still, we will provide the CLIP score and show that it does not capture the performance improvement from fine-tuning, as noted qualitatively, since CLIP is a standard metric for evaluating text-to-image models.

We avoided using the \textbf{exact} 300 prompts to identify the best checkpoint, so we used GPT-5.5 to rewrite 300 prompts derived from the originals. The prompt passed to GPT-5.5 is ``\textit{Generate one new CLIP evaluation prompt based on the prompt below. Requirements: Preserve the technical domain and subject, keep approximately the same length, use different wording, describe visually observable content only, do not add artistic styles, and return only the new prompt.}'' The ``\textit{prompt below}'' refers to the original prompt, and we used the OpenAI API to automate this for 300 captions. Table \ref{tab:props} shows three examples of original prompts and slightly modified versions by GPT-5.5. The complete list of the 300 modified prompts is available in the \textbf{Supplementary Materials} in a spreadsheet. 

\begin{table}[h!]
  \centering
  \footnotesize
  \caption{Original testing prompts and the modified ones by GPT-5.5.}
    \begin{tabular}{p{25em}p{25em}}
    \toprule
    \multicolumn{1}{l}{\textbf{Original Prompt}} & \multicolumn{1}{l}{\textbf{Modified prompt by GPT-5.5}} \\
    \midrule
    Real image of nuclear icebreaker Ural sailing in the sea. & Photograph of nuclear icebreaker Ural moving across the sea. \\
    \midrule
    ZETA generator model, the UK's first nuclear fusion experiment & A model of the ZETA generator, Britain’s first nuclear fusion experiment \\
    \midrule
    Westinghouse VVER-440 NOVA E-6 fuel assemblies at Dukovany, Czech Republic. & Westinghouse VVER-440 NOVA E-6 fuel bundles at the Dukovany nuclear facility, Czech Republic. \\
    \bottomrule
    \end{tabular}%
  \label{tab:props}%
\end{table}%

\subsection{Qualitative Strategy}
\label{subsec:quality}

\begin{figure}[t!]
    \centering
    \includegraphics[width=0.9\linewidth]{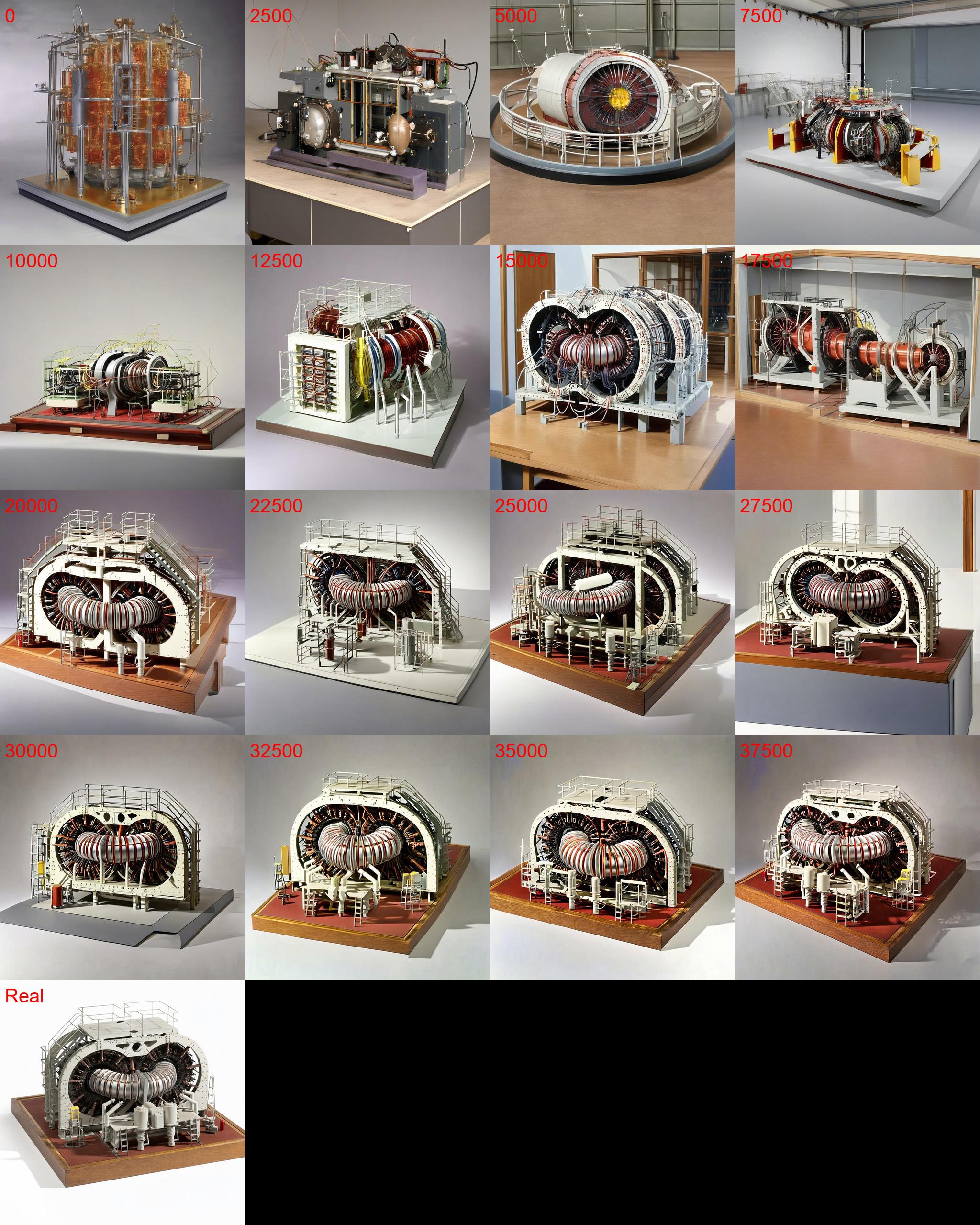}
    \caption{Images generated by SDXL for the prompt ``\textit{A model of the ZETA generator, Britain's first nuclear fusion experiment}''. The first left image in the first row is generated by the zero-shot model (0), followed by images from the 15 saved checkpoints (2500, 5000, ..., 37500). The single image in the last row is the real image.}
    \label{fig:qual_sam}
\end{figure}

The qualitative assessment of the generated images is inevitable, even if we develop metric(s) tailored for nuclear energy imagery, as such metrics must be validated. Therefore, KID and CMMD scores should be validated. Accordingly, we combined the real images with images generated at all saved checkpoints and at zero-shot, as shown in the example in Figure \ref{fig:qual_sam}. The left image in the first row is generated by the zero-shot SDXL using the prompt ``\textit{A model of the ZETA generator, Britain's first nuclear fusion experiment}'', followed by 15 images generated by the 15 checkpoints (2,500, 5,000, ..., 37,500). The image in the last row is the real one. From this example, it can be seen that SDXL is learning the new concept with large success, with the image generated by the checkpoint saved after 37,500 training steps being the closest to the real image shown in the last row. 

\begin{figure}[t!]
    \centering
    \includegraphics[width=0.9\linewidth]{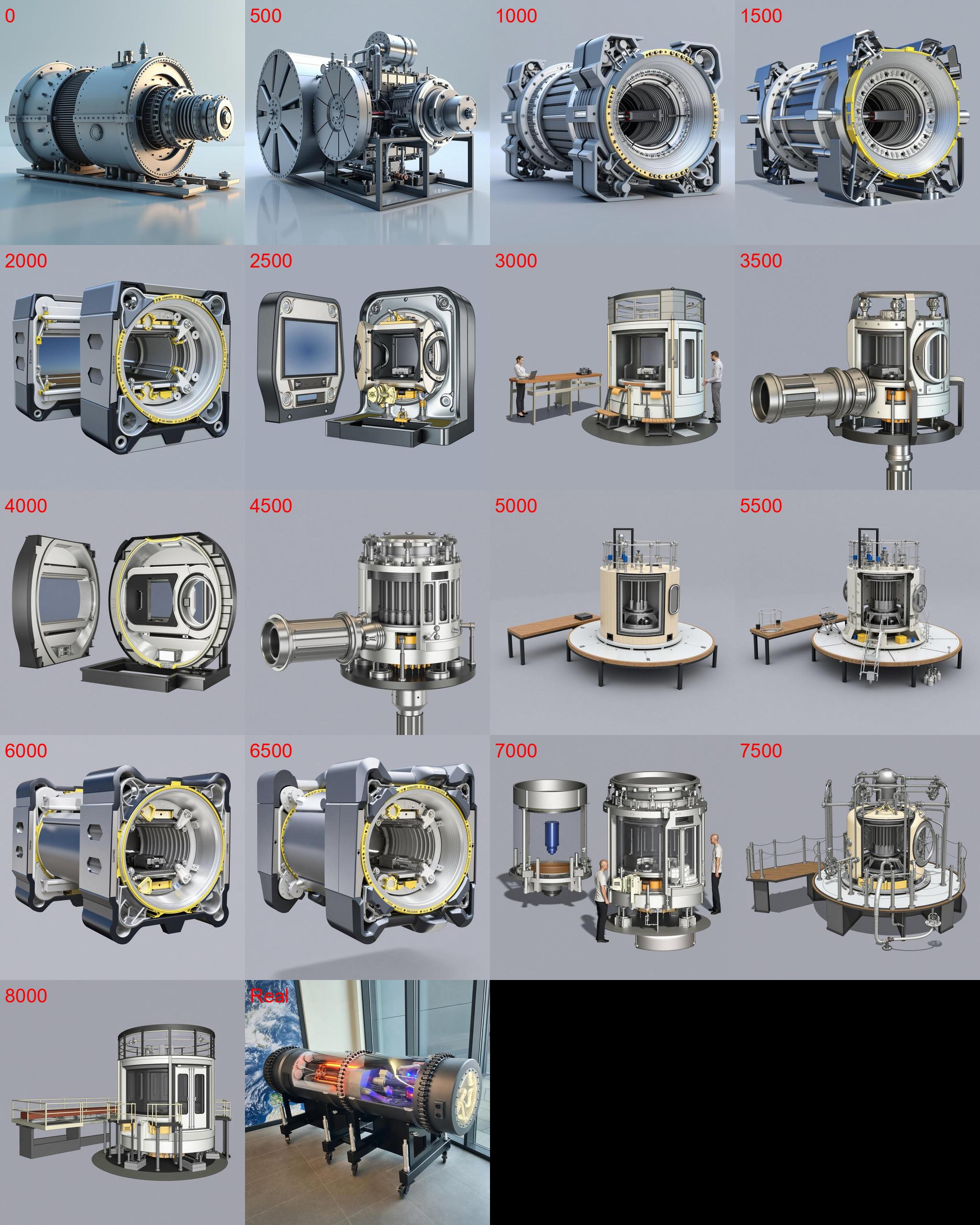}
    \caption{Images generated by Flux.1 for the prompt "\textit{Rolls-Royce compact nuclear microreactor concept image.}" The first left image in the first row is generated by the zero-shot model, followed by images generated using the 16 saved checkpoints. The second image in the last row is the real image.}
    \label{fig:qual_fail}
\end{figure}

We generated similar figures for all 300 test samples across the three models considered (SDXL, \sdvv , Flux.1), and qualitatively assessed whether the model is learning the concept like in Figure \ref{fig:qual_sam}, and if so, which model is the closest to the real images, where such a model will earn a point for every image. For example, in Figure \ref{fig:qual_sam}, the model saved after 37,500 training steps earns a point for this prompt. We will compare the model that scores the most points with the best models selected based on the quantitative metrics (CMMD, KID). We will also report the number of instances in which the model fails to learn the concept, as shown in Figure \ref{fig:qual_fail}. The prompt used is ``\textit{Rolls-Royce compact nuclear microreactor concept image.}'' For such instances, none of the checkpoints receive any points.

It should be noted that such images can be given to commercial large language models such as GPT or Gemini to determine which image is closest to the real one and automate the process, \textit{but we prefer that the qualitative process remain purely human in this study}.

\section{Results and Discussions}
\label{sec:forward_res}

\subsection{Quantitative Assessment}

\begin{figure}[h!]
    \centering
    \includegraphics[scale=0.52]{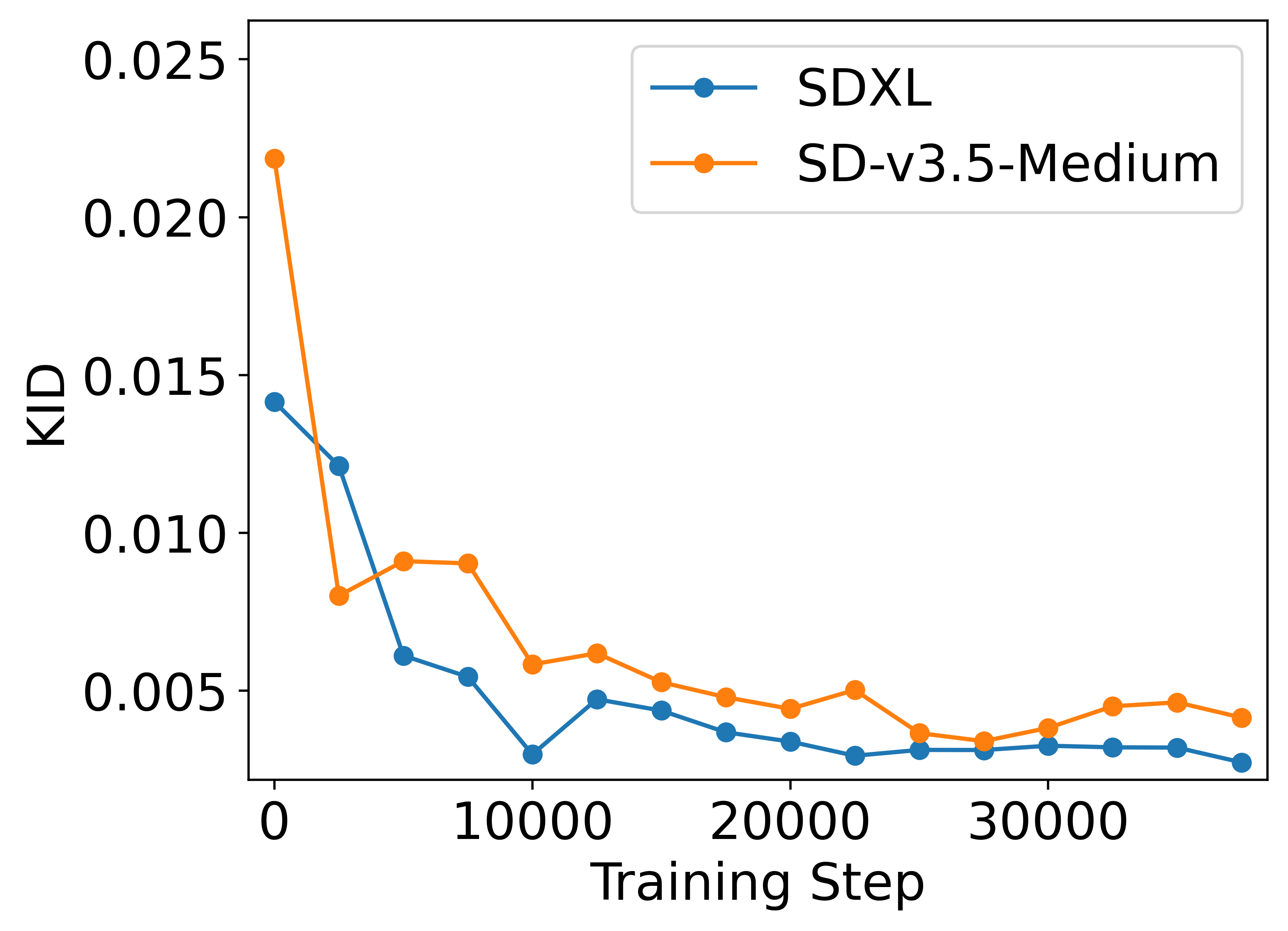} \hspace{0.15cm} \includegraphics[scale=0.52]{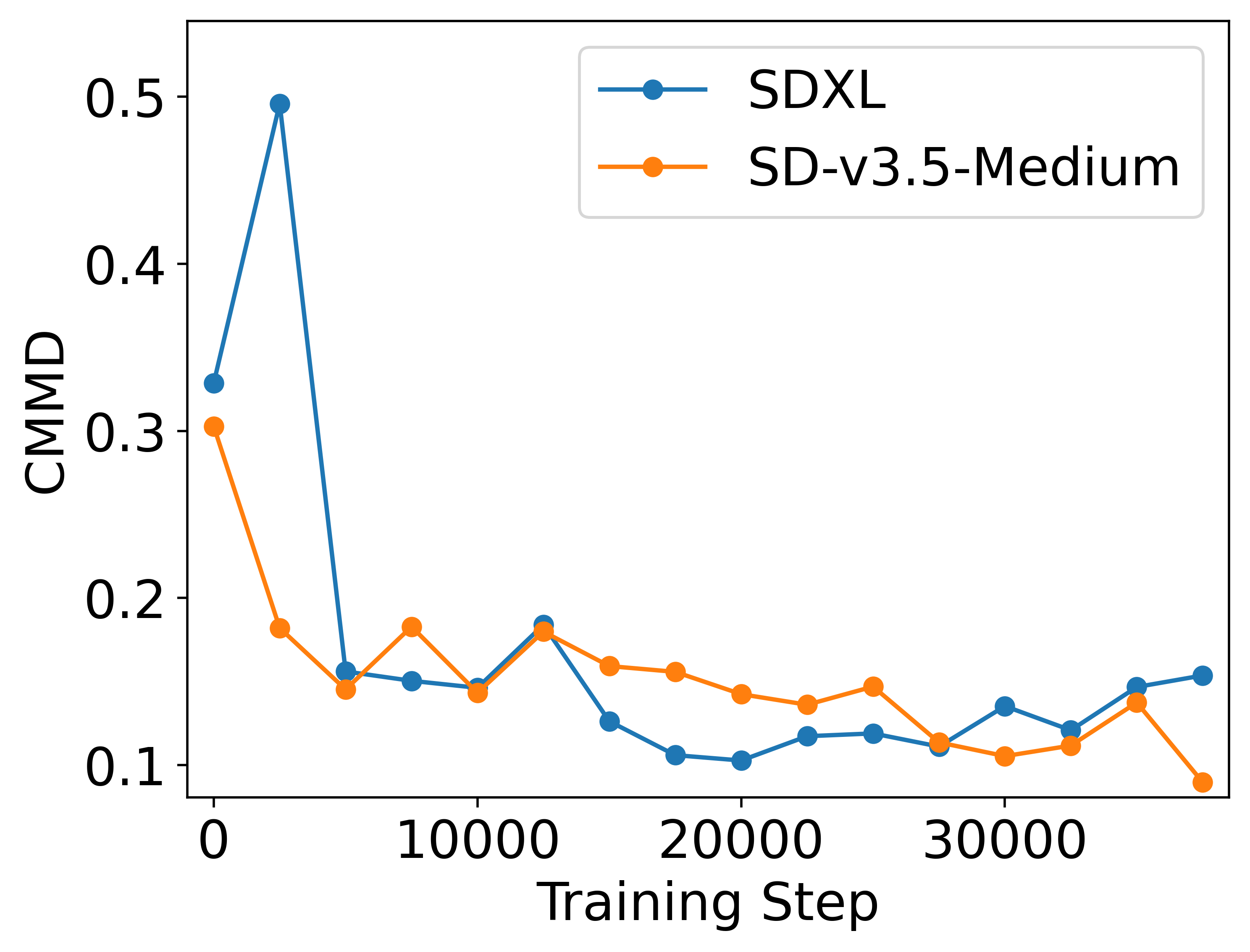} \\
    \includegraphics[scale=0.52]{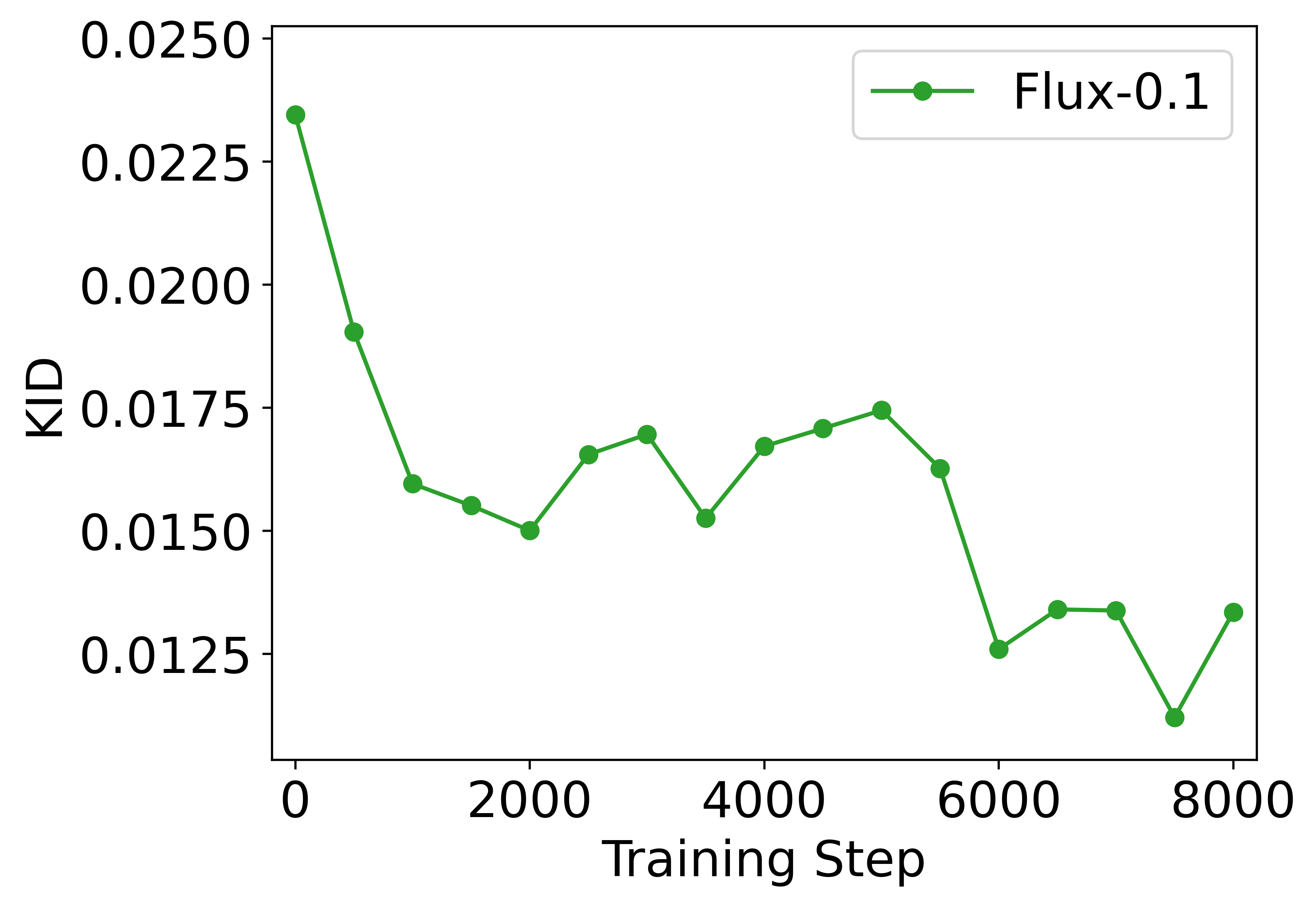} \hspace{0.25cm} \includegraphics[scale=0.52]{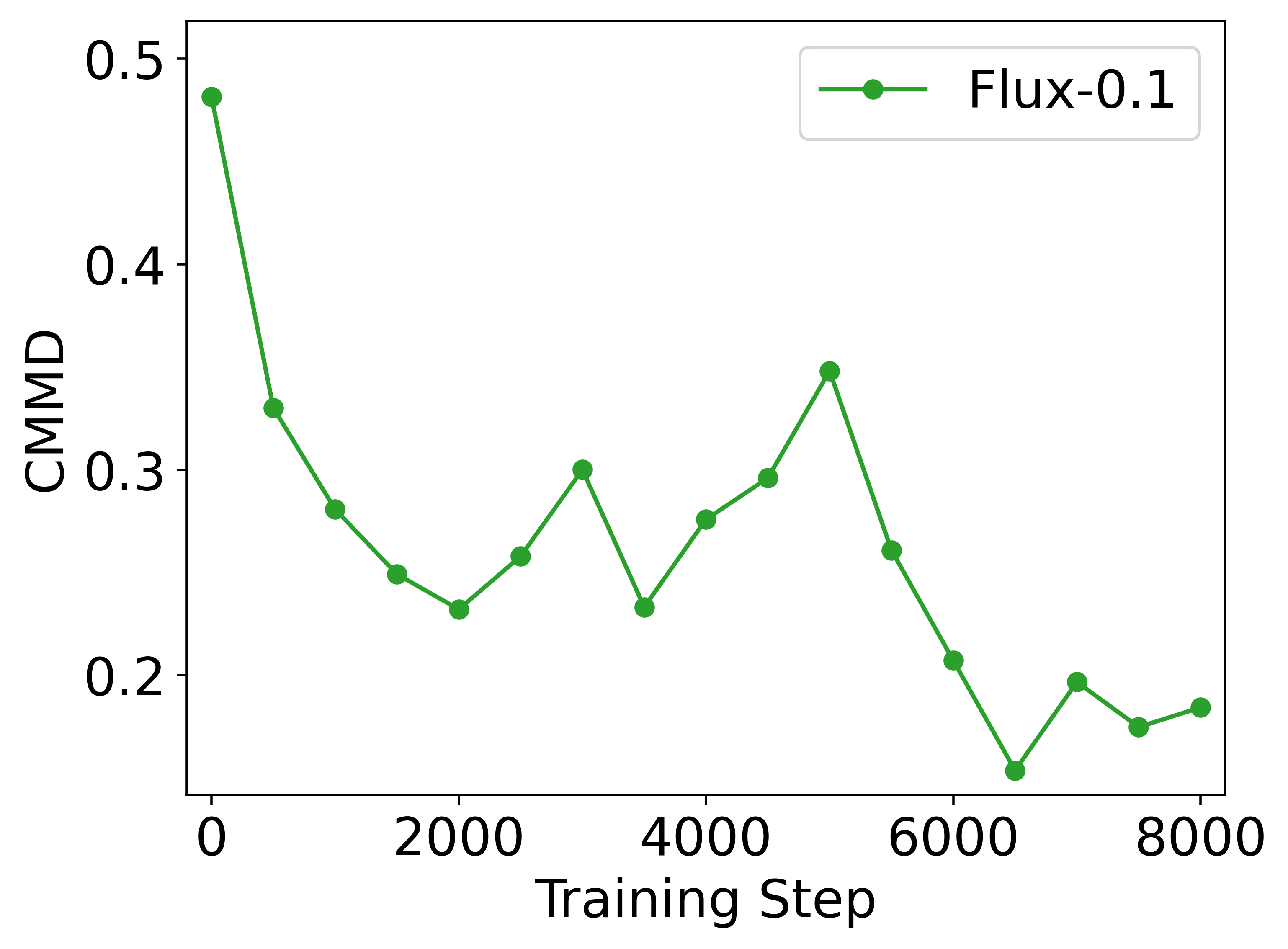} \\
    \caption{KID and CMMD for SDXL, SD-v3.5-Medium, and Flux.1}
    \label{fig:quant}
\end{figure}

Figure \ref{fig:quant} shows KID and CMMD for SDXL and \sdv, calculated over 300 images generated using the zero-shot model and 15 checkpoints saved at training steps [2500, ..., 37500]. The Figure also shows KID and CMMD for Flux.1, calculated for the zero-shot model and 16 checkpoints saved at [500, 1000, ..., 8000]. The decreasing trend in both KID and CMMD with training steps indicates greater fidelity between generated images and real images compared with the zero-shot model at training step = 0. The KID curve indicates that SDXL outperforms \sdv across all checkpoints, while the CMMD shows variation between the two models as training steps increase. The two metrics do not agree on the best checkpoint; the lowest KID and CMMD values belong to different checkpoints for both SDXL and \sdvv. For SDXL, the best (lowest) CMMD is achieved by the model saved after 20,000 training steps (CMMD=0.103), while the best (lowest) KID is achieved by the last checkpoint saved after 37,500 training steps (KID=0.003). For \sdvv, the best CMMD is achieved at the last checkpoint saved after 37,500 training steps (CMMD=0.090), while the best KID is achieved at the checkpoint saved after 27,500 training steps (KID=0.003). 

For Flux.1, the decrease in CMMD and KID shown in the lower row of Figure \ref{fig:quant} indicates improved performance with fine-tuning. The lowest CMMD and KID scores are 0.154 and 0.011, respectively, achieved after 6500 and 7500 training steps. Both scores are higher than those obtained by the SD models and indicate poorer performance. Table \ref{tab:quant_summ} summarizes the best scores achieved by the three models and the checkpoints that achieved them.

\begin{table}[h!]
  \centering
  \caption{The (lowest) CMMD and KID scores achieved by the three models and the related checkpoint saved after $n$ training steps, written in parentheses.}
    \begin{tabular}{lcc}
    \toprule
     & \multicolumn{1}{l}{Best CMMD} & \multicolumn{1}{l}{Best KID} \\
    \midrule
    SDXL            & 0.103 ($n=$ 20,000)    & 0.003 (37,500) \\
    SD-v3.5-Medium  & 0.090 (37,500)    & 0.003(27,500) \\
    Flux.1          & 0.154 (6,500)     & 0.011(7,500) \\
    \bottomrule
    \end{tabular}%
  \label{tab:quant_summ}%
\end{table}%

Figure \ref{fig:clip} in \ref{app:clip} shows the CLIP score calculated at the 15-16 checkpoints of the three considered models. The scores indicate that the zero-shot models are the best (highest CLIP score) for SDXL and \sdvv, while the best checkpoint for Flux.1 is the one saved after 1,500 training steps, with a slightly higher CLIP score than the zero-shot model. The CLIP scores indicate that fine-tuning is not effective for all three models and don't agree with CMMD and KID. More recent text-image alignment metrics, Image Reward \cite{xu2023imagereward}, and visual question-answering metrics (VQ$^2$) \cite{yarom2023you} were also tried but failed to capture the improvement due to fine-tuning. All of these observations will be validated qualitatively, including the best checkpoint for each model and the performance ranking of the three models.

\subsection{Qualitative Assessment}

\begin{figure} [h!]
    \centering
    \includegraphics[scale=0.55]{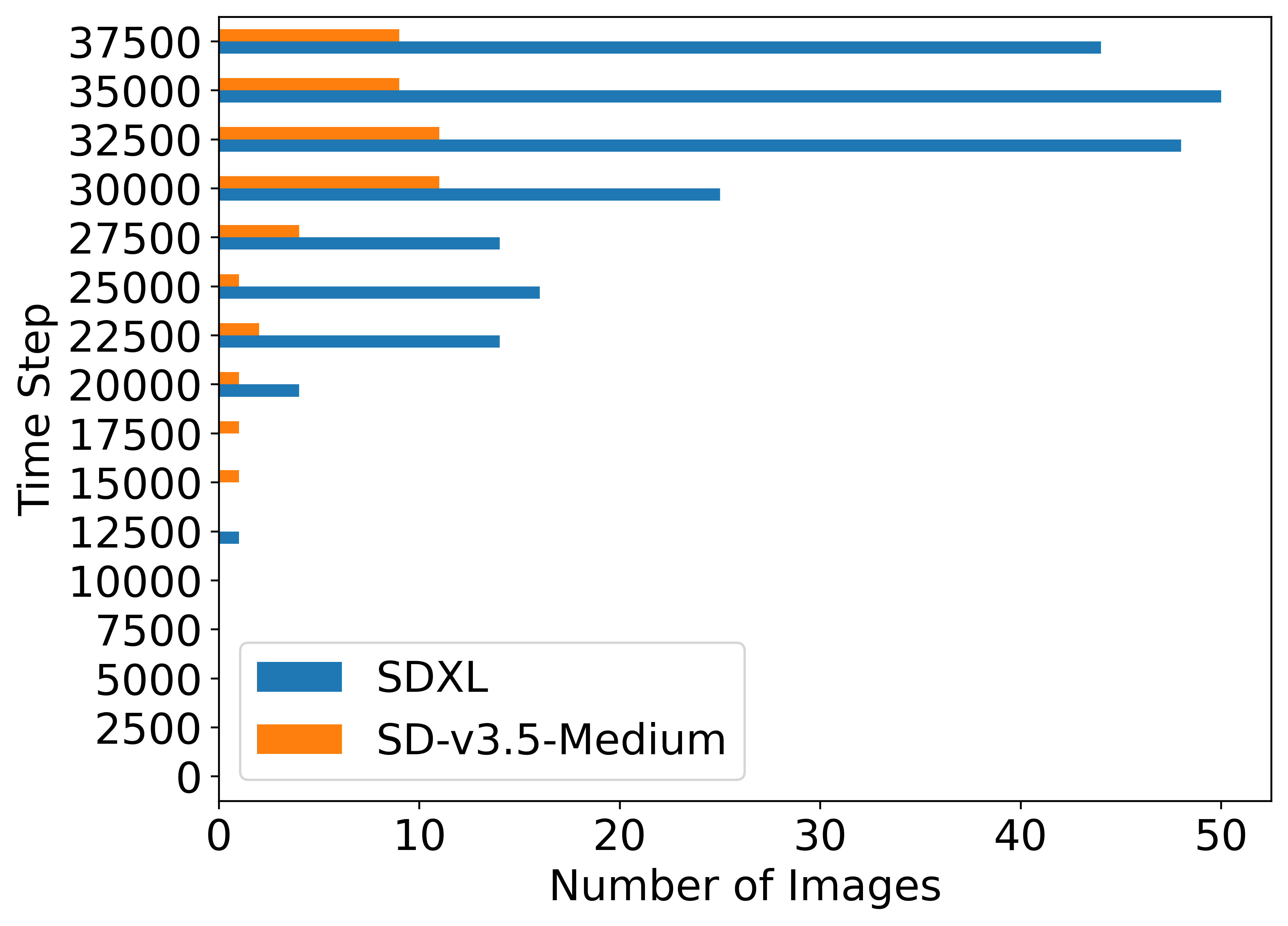} \\
    \vspace{0.5cm}
    \includegraphics[scale=0.55]{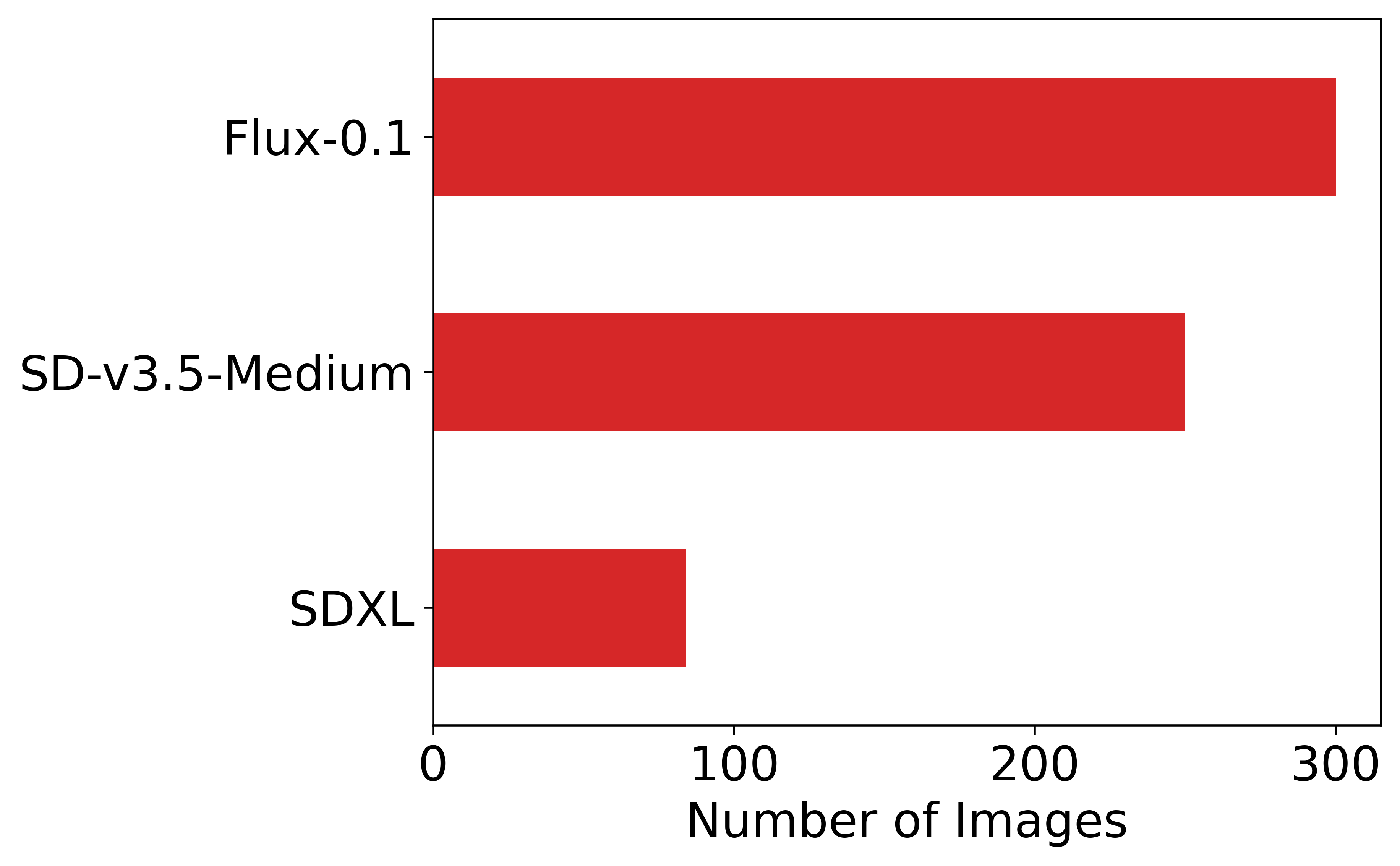}
    \caption{\textbf{Top}: The number of images that are closest to real images for each checkpoint, decided qualitatively by a human judge. \textbf{Bottom}: The number of images where none of the checkpoints generated a close image to the real one, decided qualitatively by a human judge.}
    \label{fig:quality}
\end{figure}

We explained our qualitative strategy in subsection \ref{subsec:quality}, where we grouped all the images generated by all checkpoints with the real image, as shown in Figures \ref{fig:qual_sam} and \ref{fig:qual_fail}, and a human judge decided which image was closest to the real one and gave that checkpoint a point. The top panel of Figure \ref{fig:quality} shows the number of such images for each checkpoint for both SDXL and \sdv. The bottom panel shows the number of images where none of the checkpoints generated an image close to the real image, as in the example shown in Figure \ref{fig:qual_fail}. Since Flux.1 failed in all 300 images, it was excluded from the top panel of Figure \ref{fig:quality}. 

The top panel of Figure \ref{fig:quality} confirms that fine-tuning improves the accuracy of images generated by SD models compared with the zero-shot models. However, the quantitative metrics CMMD and KID indicate comparable performance between SDXL and \sdvv, with CMMD indicating that \sdv performs better. However, the bottom panel of Figure \ref{fig:quality} shows that \sdv failed to generate images reasonably close to the real ones in 250 instances, while SDXL failed in 84 instances. Accordingly, SDXL exhibits significantly better performance than \sdvv. While the images are not accurate, other subtle details such as lighting, colors, and the place may explain the improvement in CMMD and KID for \sdv and Flux.1, as shown in Figure \ref{fig:quant}. The number of trainable parameters in SDXL is larger (2.77 billion) than \sdvv (2.5 billion), but this slight difference may not be the main reason for SDXL's superior performance.

The top three SDXL checkpoints are those saved at 35,000 (50 points), 32,500 (48 points), and 37,500 (44 points). Compared with Table \ref{tab:quant_summ}, KID choice looks more reasonable for the best checkpoint (37,500) than CMMD (20,000), where the checkpoint saved after 20,000 training steps got only 4 points. Since the best checkpoints are the last three, SDXL may benefit from longer training. For \sdvv, the best models got only 11 points (30,000, 32,500), followed by the last two checkpoints (35,000, 37,500) with only 9 points. In contrast to SDXL, CMMD selects the best model (37,500) better than KID (27,500). Similar to SDXL, \sdv may benefit from longer training, but longer training can also lead to overfitting, resulting in less accurate images compared with real ones.

The qualitative assessment of the images generated by Flux.1 showed that the model is not learning and is using the wrong concept and building on it. This is probably due to LoRA compression, which significantly reduces the number of trainable parameters compared with the full model, which has 12 billion parameters. However, another possibility is that the model is overfitting due to the use of a high LoRA rank (256$\times$256). Accordingly, we tried fine-tuning with lower ranks (128$\times$128, 16$\times$16) and observed images generated for 7 validation prompts every 500 steps, but they were similar to those produced during training at (256$\times$256).

The qualitative assessment also disagrees with the CLIP scores shown in Figure \ref{fig:clip}, which indicate that the zero-shot SDXL and \sdv models are the best and that fine-tuning is completely ineffective. As mentioned in subsection \ref{subsec:quant}, calculating the CLIP score requires an image embedding model that may fail on unfamiliar images, such as nuclear energy images. It should be noted that CMMD and KID also use image embedding models, but the availability of a reference image for comparison explains why these metrics perform better than CLIP. Still, KID and CMMD would benefit from an image-embedding model trained on nuclear-energy images.

\begin{figure}[t!]
    \centering
    \includegraphics[width=\linewidth]{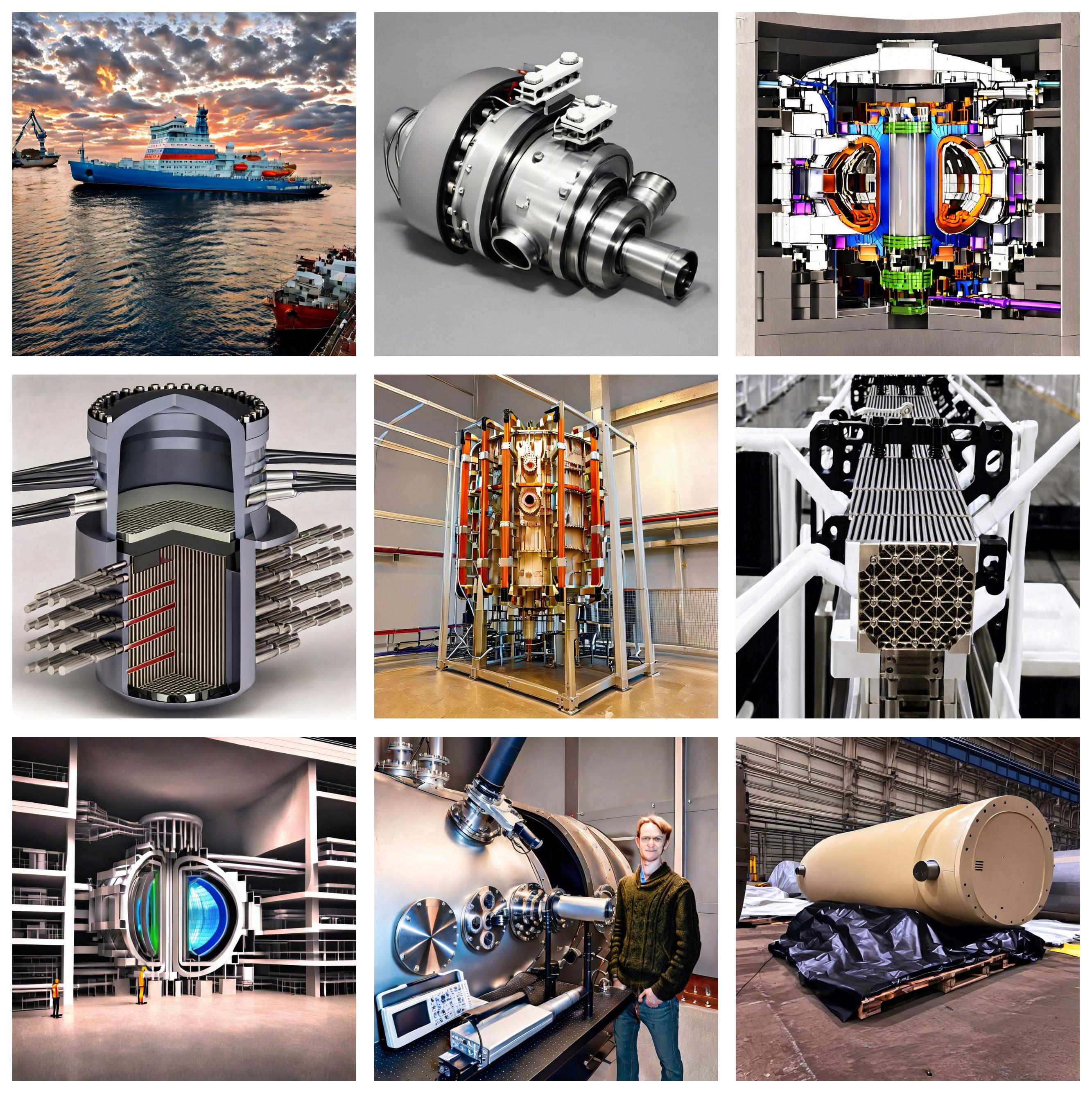}
    \caption{Nine examples where SDXL checkpoints generated very good-quality images compared with the real ones. Starting from top left, the images represent:  (1) nuclear icebreaker Ural, (2) Mini-Brayton Rotating Unit, (3) ITER tokamak schematic, (4) Canadian supercritical water-cooled reactor core schematic, (5) SMART spherical-tokamak fusion device, (6) VEL TVS-K fuel assembly, (7) General Atomics' steady-state advanced-tokamak fusion concept, (8) Light Fusion CEO beside the 22-m gas gun, and (9) Holtec's HI-STAR 149 transport cask.}
    \label{fig:imgs_mat}
\end{figure}

Overall, the qualitative assessment remains the standard for evaluating the quality of generated images, selecting the best checkpoint, and even ranking multiple models. Figure \ref{fig:quant} shows a slow decrease in CMMD and KID after 10,000 training steps for SD models. However, the qualitative assessment in Figure \ref{fig:quality} shows that the models trained for up to 20,000 steps are unusable. The qualitative assessment will automatically identify the best model among multiple text-to-image generative models. Figure \ref{fig:imgs_mat} shows nine additional examples where SDXL generated very good images.

\subsection{Quality of Generating Multiple Images and Prompting Effect}

\begin{figure}[h!]
    \centering
    \includegraphics[width=\linewidth]{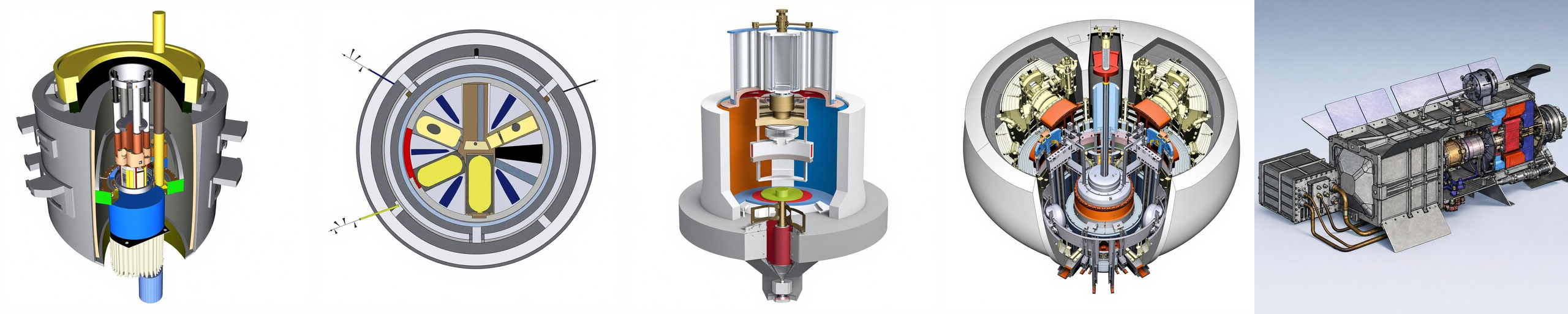}
    \caption{From left to right: 4 Images generated by the last Flux.1 checkpoint for the prompt ``\textit{Cross-section of the Advanced Stirling Radioisotope Generator (ASRG).}'' and the real image is the last image on the right.}
    \label{fig:multi}
\end{figure}

The text-to-image models can generate different images for the same prompt. Therefore, we selected 20 prompts from the pool for which all three models failed to produce an image reasonably close to the real one, and generated 4 additional images from those produced in the last two subsections to assess whether generating multiple images would yield an accurate one(s). For SD models, we selected the checkpoints with the best qualitative performance (SDXL, 35,000), (\sdvv, 32,500). For Flux.1, we do not have an optimal checkpoint, so we picked the last checkpoint. Figure \ref{fig:multi} shows 4 images generated by Flux.1 starting from left to right, and the fifth image on the right is the real image. The prompt used is ``\textit{Cross-section of the Advanced Stirling Radioisotope Generator (ASRG)}''.  It can be seen that the images remain inaccurate for this sample; this was the case for all 40 instances distributed among Flux.1 and \sdvv. For SDXL, we observed images that were closer to the real ones in only 3 of 20 instances, and even then, they were still not quite accurate. For SDXL, we also used the SDXL image refiner model in the inference. The findings suggest that the overall quality of multiple images still depends on the quality of fine-tuning.

\begin{figure} [t!]
    \centering
    \includegraphics[width=0.75\linewidth]{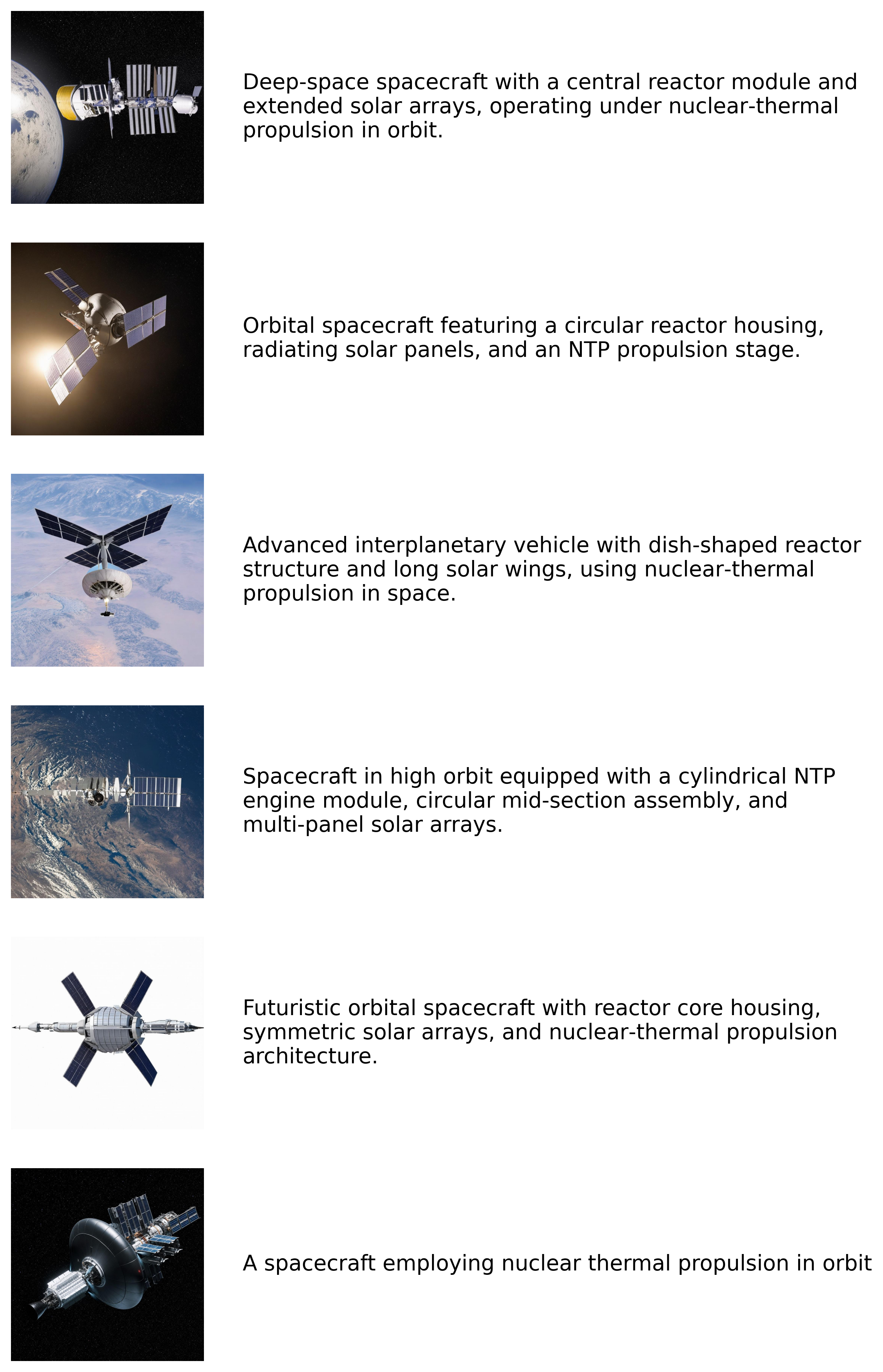}
    \caption{Images generated (left) for five prompts (right) extracted from the original prompt and image shown in the last row. The images were generated by \sdv checkpoint saved after 32,500 training steps.}
    \label{fig:mp}
\end{figure}

To assess the effect of prompting, we selected 5 prompts that all three models failed to generate accurate images for, and generated 5 variations for each prompt to evaluate the impact of word selection. Figure \ref{fig:mp} shows five images generated by \sdv using five prompts extracted from the original prompt and image shown in the last row. The original image shows a concept of a spacecraft powered by nuclear thermal propulsion. It can be seen that changing the prompt alters the generated image, but the spacecraft concept is still inaccurate. This is the case for the other four prompts and also the other two models (SDXL, Flux.1). As a result, as with generating multiple images, accurate generation depends mainly on accurate fine-tuning and re-prompting did not help with the performance. 

\section{Performance of Commercial Models}
In this section, we compare quantitatively and qualitatively the performance of our fine-tuned models with three common text-to-image commercial models: 
\begin{itemize}
    \item \textbf{GPT-Image-2.0}: This is the latest text-to-image model by OpenAI. We used the OpenAI Python API to generate 300 images using the 300 testing prompts. The cost was around \$64 to generate 300 high-quality 1024$\times$1024 images. The inference time for the 300 images was the longest, at around 16 hours, with no generation failures.
    
    \item \textbf{Gemini-3.1-Flash-Image}: This is Google's latest model. We used the Python API to generate the 300 images, where the cost was around \$20. While the inference time is dominantly lower than that of gpt-image-2.0 (around 45 minutes), Gemini-3.1-Flash-Image failed to generate images for 15 prompts. According to Google Studio, these are due to ``internal server errors'' or ``service unavailable''. Keeping only failed prompts and rerunning inference multiple times gradually reduced failures to zero. 
    
    \item \textbf{Midjourney}: The developers of this model do not provide an API for generating a large number of images. Therefore, we had to use a third-party application, LegNext AI \cite{Legnext}, which is dedicated to Midjourney and provides a reliable API. LegNext AI's Pro subscription costs \$30 per month and gives credit to generate 375 images. The API successfully generated 297 images and failed on 3, which we manually generated using the LegNext interface. Midjourney generates four images per prompt, so we randomly selected one for each prompt; the inference time was around 7 hours.
\end{itemize}

The standard approach should be to qualitatively assess the three models' generations and determine whether the concepts are generally correct. Using CMMD and KID, we will compare the generated images with real images that the models did not see. As a result, these metrics may indicate that the commercial models have comparable or even worse performance than the zero-shot open-source models, as shown in Table \ref{tab:comm_met} in \ref{app:comm_quant}.

\begin{figure} [h]
    \centering
    \includegraphics[width=\linewidth]{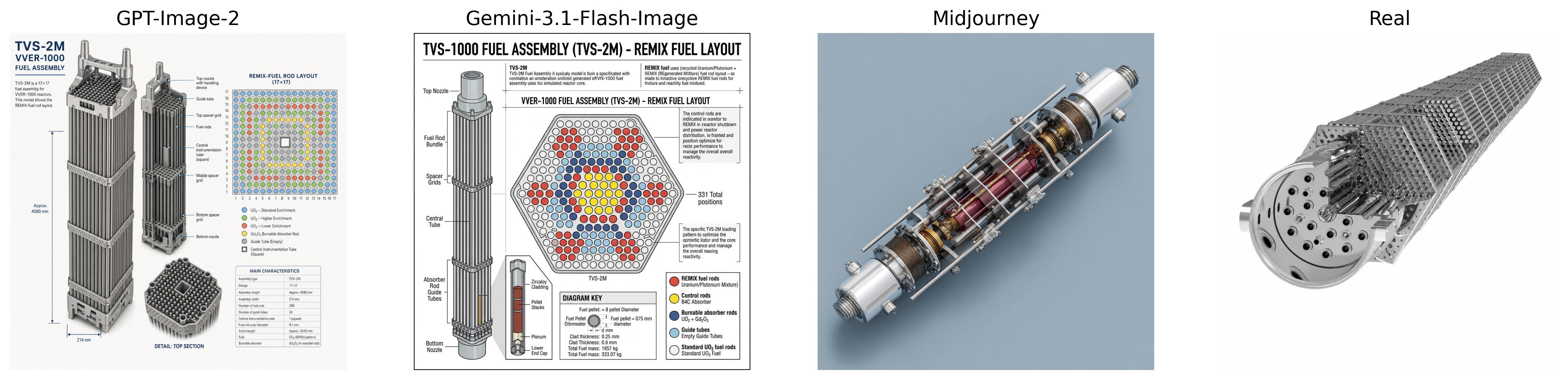}
    \caption{Images generated by the three commercial models considered and the real image for the prompt: "\textit{TVS-2M VVER-1000 fuel assembly model showing a REMIX-fuel rod layout.}"}
    \label{fig:qual_comm}
\end{figure}

We combined the generations of the three commercial models with the real image, as shown in Figure \ref{fig:qual_comm}. The prompt used is ``\textit{TVS-2M VVER-1000 fuel assembly model showing a REMIX-fuel rod layout.}''. The figure shows that GPT-Image-2 and Gemini-3.1-Flash-Image recognize the concept of a fuel assembly, whereas Midjourney does not. However, the exact design details for the TVS-2M VVER-1000 fuel assembly are not accurate. Therefore, we noticed that the GPT and Gemini models are significantly better than Midjourney at nuclear energy concepts across nearly all 300 images, but still do not capture the details accurately. Also, all three models failed in other instances to generate accurate images for some concepts or to produce the correct views of nuclear reactors in different places around the globe. We also observed rich use of annotations in the images generated by GPT-Image-2 and Gemini-3.1-Flash-Image. The English words used are correct but can be used inaccurately to describe something in the image. For example, the word ``compressor'' is pointing to a ``pump'' symbol in a thermal cycle diagram generated by Gemini-3.1-Flash-Image. As mentioned before, we noticed that the open-source Stable Diffusion and Flux.1 models were unable to annotate images with correct English words; therefore, we removed the text from the images used for fine-tuning.

Overall, we cannot rely on commercial models to provide images related to nuclear energy, despite their attractive generation. Such models can provide illustrative images of general concepts in nuclear energy, but we still cannot rely on them to generate creative designs that humans might not think of, or to create accurate videos of a tour in a research reactor at a lab or university. This implies the need to fine-tune models tailored for nuclear energy that will likely outperform the commercial models. The small SDXL model showed promising performance in learning the concept but still cannot capture all the details and cannot annotate with proper English words. Larger models still need to be fine-tuned with more images and variations where they struggle to learn the concept. The limited hardware capacity currently prevents us from pursuing this direction, which we are planning to resolve in our future study that builds on this work.

\section{Concluding Remarks and Limitations}
\label{sec:conc}

In this study, we considered fine-tuning text-to-image generative AI models to produce images related to nuclear energy. Accordingly, we collected 1000 captioned images from handbooks, encyclopedias, publications, and nuclear energy news websites. We fine-tuned the full Stable Diffusion XL and v-3.5-Medium models, as well as LoRA Flux.1. The quantitative assessment using 300 testing prompts by KID and CMMD indicated significant performance improvements over the zero-shot models for all three models, with SDXL and \sdv showing comparable performance and outperforming Flux.1. Nevertheless, the qualitative assessment showed that SDXL is significantly better than \sdv, and Flux.1 failed on all 300 test prompts. Findings also showed that both generating multiple images and varying the input prompt do not guarantee an accurate image. We also assessed images generated by three commercial text-to-image models: GPT-Image-2, Gemini-3.1-Flash-Image, and Midjourney, and observed accurate images from GPT-Image-2 and Gemini-3.1-Flash-Image for general concepts of nuclear energy, while Midjourney generations were inaccurate. However, as the prompts became specific, all three failed to generate accurate images, suggesting the necessity for fine-tuning. 

Among the limitations of this work are the cost of human involvement in assessing generations and the need for a reference image for quantitative assessment. The text-image alignment metric CLIP did not provide an accurate assessment consistent with the qualitative assessment. We also built a dataset of a wide variety of nuclear energy images and used the captions provided by the sources, which may be short and not sufficiently descriptive. Future research directions include increasing the dataset size to hundreds of thousands through augmentation and generating variations of the real images. We will need this to develop a metric tailored to assess nuclear energy images, using an image embedding model trained on a large dataset of thousands of images. Such a metric will help assess the model's creativity and rank the multiple images generated. We will also consider fine-tuning larger models, such as Stable Diffusion v-3.5-Large, Ideogram 4.0, and the full Flux.1 model to capture the details that SDXL could not, through expanded GPU hardware that the authors will acquire in the future. 

\section*{Data Availability}
The image dataset can be sent upon reasonable request. The best SDXL checkpoint can be used to generate 300 images using the 300 testing prompts from this GitHub repository \url{https://github.com/aims-umich/NE_text_image.git}.

\section*{Acknowledgment}
This work was sponsored by the Department of Energy Office of Nuclear Energy through the Nuclear Energy University Programs (Award: DE-NE0009497). Furthermore, the first author (M. I. Radaideh) was partially supported by the Michigan Institute for Computational Discovery and Engineering (MICDE) Research Scholar Program during this research work.

\appendix
\renewcommand\thefigure{\thesection.\arabic{figure}}   
\setcounter{figure}{1}
\counterwithin{figure}{section}
\counterwithin{table}{section}

\section{Prompt used for GPT Models to Filter Unwanted Images from PDFs}
\label{sec:appA}

\begin{tcolorbox}[title=GPT-4.1-nano and GPT-5-nano Prompt, breakable]
You are a precision-focused expert in analyzing technical and scientific documents. Your task is to perform a strict two-stage classification for EACH of the provided images. \\

**Your output MUST be a valid JSON object where the keys are the exact filenames of the images provided, and the values are objects containing three keys: "is\_data\_plot", "is\_nuclear\_schematic", and "reasoning".** \\

Example Output Structure: \newline
\{ \newline
  "image1.jpg": \{ \newline
    "is\_data\_plot": false, \newline
    "is\_nuclear\_schematic": true, \newline
    "reasoning": "This image shows a reactor core layout..." \newline
  \}, \newline
  "image2.png": \{ \newline
    "is\_data\_plot": true, \newline
    "is\_nuclear\_schematic": false, \newline
    "reasoning": "This is a line graph showing neutron flux..." \newline
  \} \newline
\} \\

**Stage 1: Data Plot Identification** \newline
First, determine if the image is a 'data plot'. A data plot's primary purpose is to graphically represent numerical data within a coordinate system. \\

Key features of a 'data plot' include: \newline
- X-Y or other axes with labeled tick marks. \newline
- Data shown as lines, curves, points, bars, or surfaces. \newline
- A legend, title, or annotations explaining the data. \newline
- Examples: line graphs, bar charts, scatter plots, histograms. \\

If the image uses a coordinate system to show a relationship between variables, it IS a data plot. \\

**Stage 2: Nuclear Schematic Identification** \newline
Second, determine if the image is a 'nuclear schematic'. A nuclear schematic is a technical diagram illustrating the physical structure, layout, or components of a nuclear reactor or related systems. \\

Key features of a 'nuclear schematic' include: \newline
- Depictions of reactor cores, often with hexagonal or grid-patterned fuel assemblies. \newline
- Cross-sections of reactor pressure vessels showing internal structures. \newline
- Diagrams of cooling systems, control rods, fuel pellets, or particle detectors. \newline
- Technical blueprints, 3D CAD renderings, or simulation geometry from codes like MCNP or Serpent. \newline
- Annotations and labels pointing to specific engineering components. \\

The image should be a technical illustration, not a generic photograph of equipment. \\

**Classification Schema:** \newline
- 'is\_data\_plot' (boolean): 'true' if it is a data plot, 'false' otherwise. \newline
- 'is\_nuclear\_schematic' (boolean): 'true' if it is a specific, technical nuclear schematic, 'false' otherwise. \newline
- 'reasoning' (string): A brief, one-sentence explanation for your decisions. \\

Example: An image of a reactor core layout is NOT a plot and IS a schematic. A graph of neutron flux vs. time IS a plot and IS NOT a schematic.
\end{tcolorbox}

\section{CLIP Score}
\label{app:clip}
\renewcommand{\thefigure}{B.\arabic{figure}}

\begin{figure}[h!]
    \centering
    \includegraphics[scale=0.55]{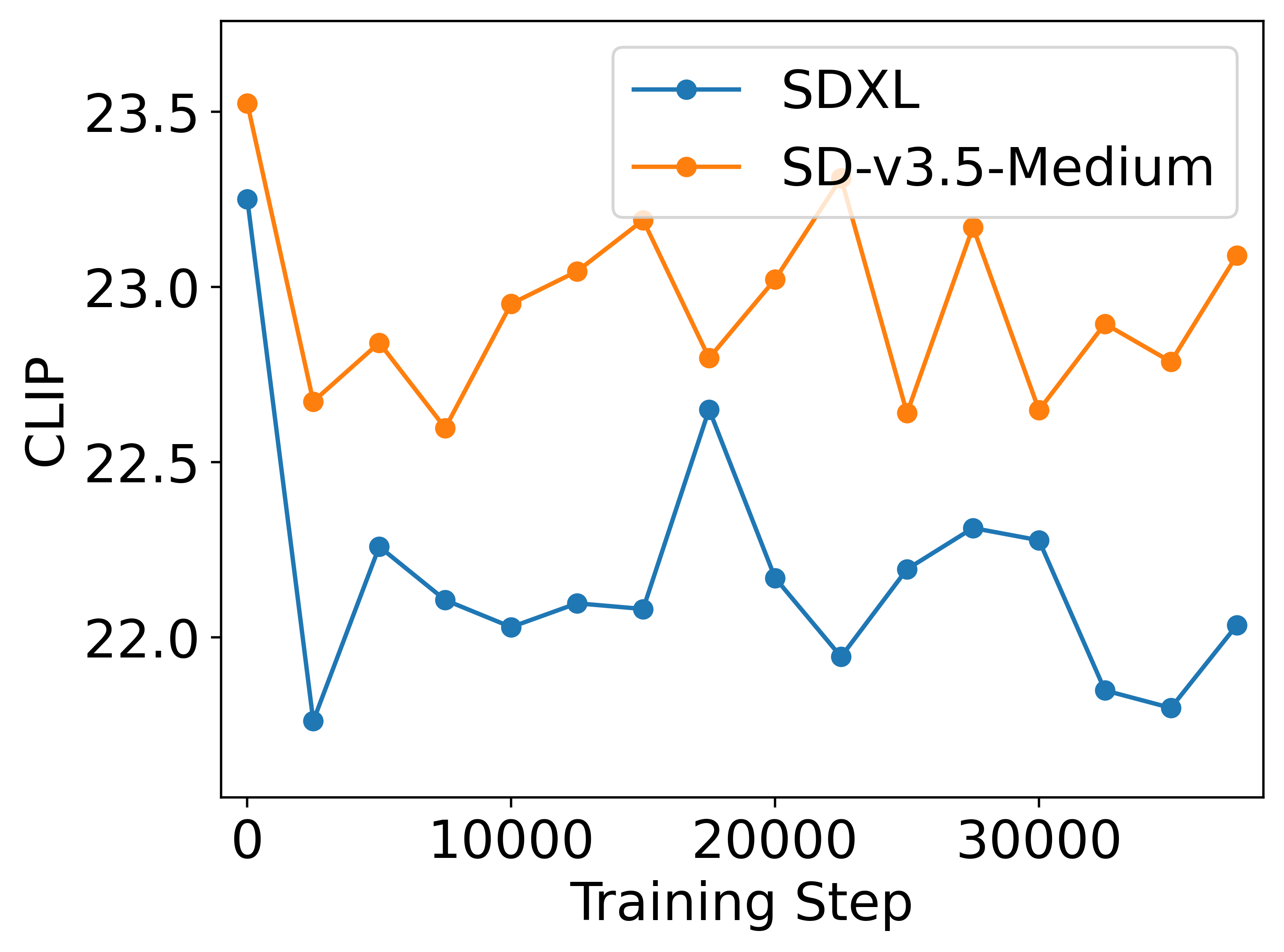} \hspace{0.15cm} \includegraphics[scale=0.55]{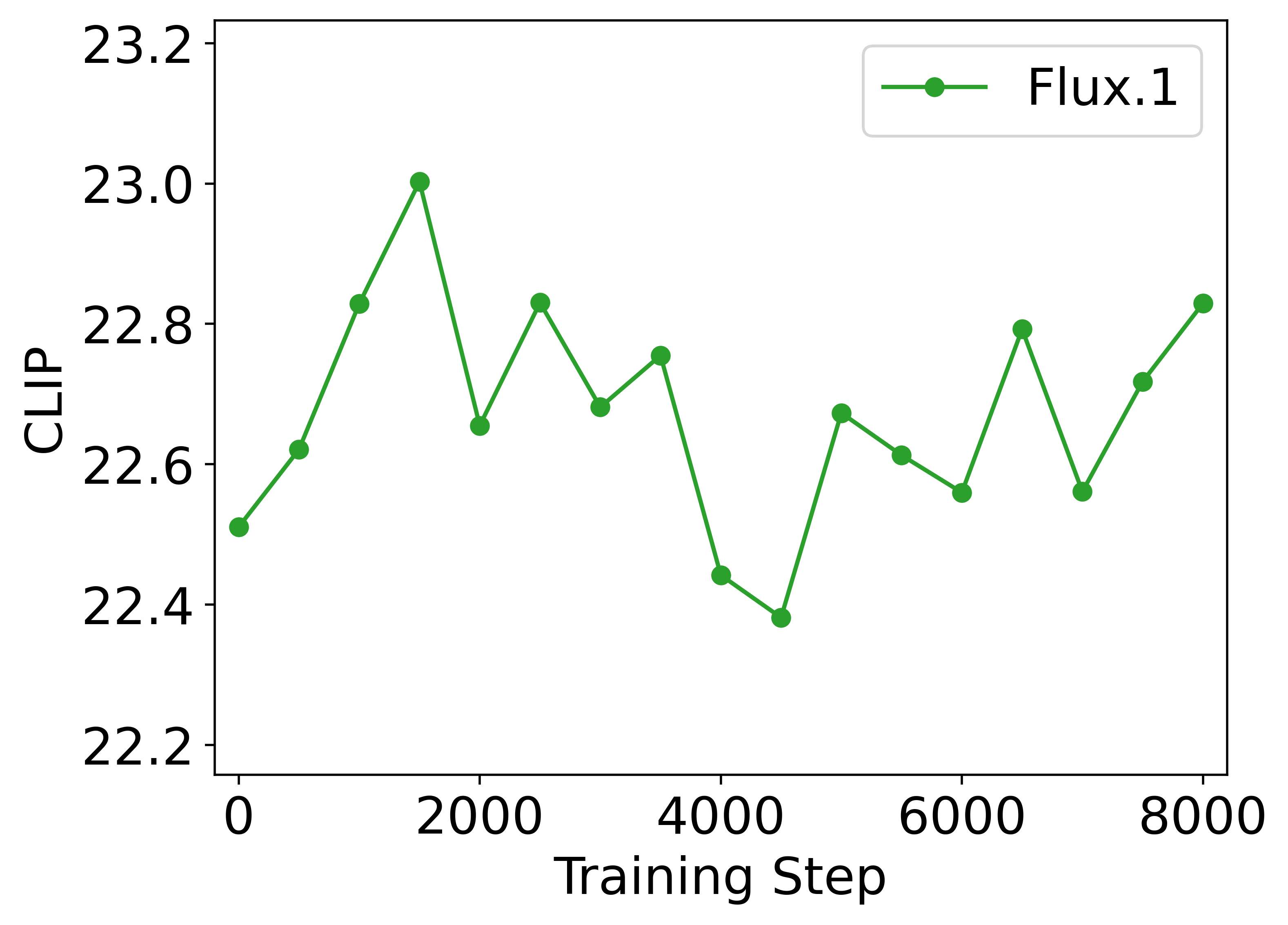} 
    \caption{Text-image alignment metric CLIP for the three considered models.}
    \label{fig:clip}
\end{figure}

Figure \ref{fig:clip} shows the CLIP score calculated for checkpoints saved every 2,500 steps for SDXL and \sdvv, and 500 for Flux.1. The CLIP score was inconclusive in this study and indicates that fine-tuning is not effective for all three models and does not agree with the conclusions seen by CMMD and KID metrics. 

\section{Quantitative Metrics for Commercial text-to-image Models}
\label{app:comm_quant}
\renewcommand{\thetable}{C.\arabic{table}}

Table \ref{tab:comm_met} shows CMMD and KID for the three commercial models considered compared with the Zero-Shot SDXL, \sdvv, and Flux.1.
\begin{table}[h!]
  \centering
  \footnotesize
  \caption{CMMD and KID for the commercial model (GPT-Image-2.0, Gemini-3.1-Flash-Image, Midjourney), and open-source zero-shot SD models and Flux.}
    \begin{tabular}{lcc}
    \toprule
    \multicolumn{1}{c}{\textbf{Model}} & \textbf{CMMD} & \textbf{KID} \\
    \midrule    
    GPT-Image-2.0                      & 0.930         & 0.024 \\
    Gemini-3.1-Flash-Image             & 0.580         & 0.018 \\
    Midjourney                         & 0.505         & 0.020 \\
    \midrule
    Zero-Shot SDXL                     & 0.329         & 0.014 \\
    Zero-Shot SD-v3.5-Medium           & 0.316         & 0.022 \\
    Zero-Shot Flux.1-dev             & 0.501         & 0.024 \\
    \bottomrule
    \end{tabular}%
  \label{tab:comm_met}%
\end{table}%

% \section{}
% \label{App:f}
% \renewcommand{\thefigure}{A.\arabic{figure}}
% \renewcommand{\thetable}{A.\arabic{table}}
% \renewcommand{\thesubsection}{A.\arabic{subsection}}
% \renewcommand{\thesubsubsection}{A.\arabic{subsection}.\arabic{subsubsection}}

%\appendix

%% References

%\bibliographystyle{elsarticle-harv}
%\bibliographystyle{elsarticle-num}
% \bibliographystyle{elsarticle-harv}
% \bibliographystyle{elsarticle-num-names}
% \bibliographystyle{model1a-num-names}
% \bibliographystyle{model1b-num-names}
% \bibliographystyle{model1c-num-names}
\small
 \bibliographystyle{model1-num-names.bst}

\bibliography{References}

\end{document}